%% file: main.tex
\documentclass[10pt,twocolumn,letterpaper]{article}

\usepackage{cvpr}      % To produce the REVIEW version
\usepackage[accsupp]{axessibility}

\input{preamble}

\definecolor{cvprblue}{rgb}{0.21,0.49,0.74}
\usepackage[pagebackref,breaklinks,colorlinks,citecolor=cvprblue]{hyperref}

\def\paperID{8638} % *** Enter the Paper ID here
\def\confName{CVPR}
\def\confYear{2024}

\title{Leak and Learn: An Attacker's Cookbook to Train Using Leaked Data from Federated Learning}
\author{Joshua C. Zhao, Ahaan Dabholkar, Atul Sharma, Saurabh Bagchi\\
Purdue University\\
{\tt\small \{zhao1207,adabholk,sharm438,sbagchi\}@purdue.edu}
}

\begin{document}
\maketitle
\input{macros}

\input{sec/0_abstract}   
\input{sec/1_intro}

\input{sec/2_related}
\input{sec/3_methodology}
\input{sec/4_experiments}
\input{sec/5_discussions}
\input{sec/6_conclusions}
\smallbreak
\input{sec/7_acknowledgements}
\smallbreak
{
    \small
    \bibliographystyle{ieeenat_fullname}
    \bibliography{main}
}

\input{sec/X_suppl}

\end{document}

%% file: preamble.tex
\usepackage[dvipsnames]{xcolor}

\usepackage{comment}
\usepackage{graphicx}
\usepackage{multirow}

%% file: macros.tex
\newcommand{\name}{\textsc{Loki}\xspace}
\newcommand{\namemeaning}{In Norse mythology, \name is a cunning trickster who had the ability to change his shape. Our attack can figuratively change its shape for different clients and trick them into giving up their private data.}

\newcommand{\saurabh}[1]{\iffalse{\textcolor{red}{Saurabh: #1}}\fi}
\newcommand{\joshua}[1]{\iffalse{\textcolor{PineGreen}{Joshua: #1}}\fi}

\newcommand{\ahaan}[1]{\iftrue{\textcolor{blue}{Ahaan: #1}}\fi}
\newcommand{\atul}[1]{\iftrue{\textcolor{purple}{Atul: #1}}\fi}

%% file: sec/0_abstract.tex
\begin{abstract}
Federated learning is a decentralized learning paradigm introduced to preserve privacy of client data. Despite this, prior work has shown that an attacker at the server can still reconstruct the private training data using only the client updates. These attacks are known as data reconstruction attacks and fall into two major categories: gradient inversion (GI) and linear layer leakage attacks (LLL). However, despite demonstrating the effectiveness of these attacks in breaching privacy, prior work has not investigated the usefulness of the reconstructed data for downstream tasks. In this work, we explore data reconstruction attacks through the lens of training and improving models with leaked data. We demonstrate the effectiveness of both GI and LLL attacks in maliciously training models using the leaked data more accurately than a benign federated learning strategy. Counter-intuitively, this bump in training quality can occur despite limited reconstruction quality or a small total number of leaked images. Finally, we show the limitations of these attacks for downstream training, individually for GI attacks and for LLL attacks. 
% Finally, we show the limitations of these attacks in computation, label restoration, and reconstruction quality through the lens of downstream task usefulness. 
% \joshua{Should we also add the GI acronym?}

% The ABSTRACT is to be in fully justified italicized text, at the top of the left-hand column, below the author and affiliation information.
% Use the word ``Abstract'' as the title, in 12-point Times, boldface type, centered relative to the column, initially capitalized.
% The abstract is to be in 10-point, single-spaced type.
% Leave two blank lines after the Abstract, then begin the main text.
% Look at previous \confName abstracts to get a feel for style and length.
\end{abstract}

%% file: sec/1_intro.tex
\begin{figure}[!t]
\begin{center}
\includegraphics[width=0.9\columnwidth]{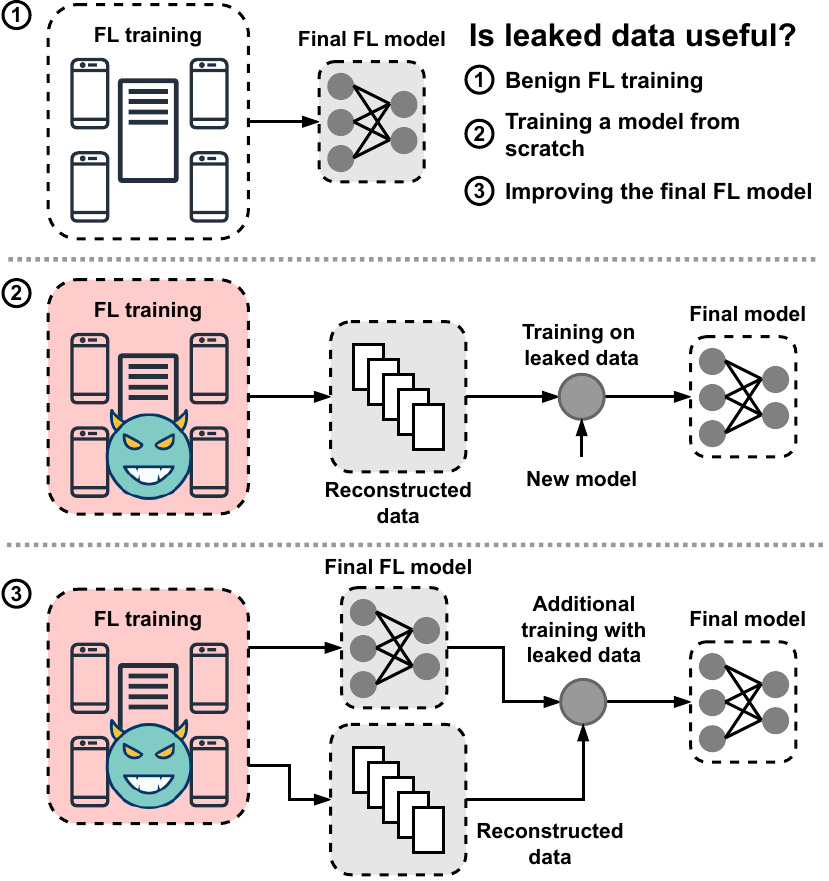}
\end{center}
\vspace*{-6mm}
\caption{\label{fig:train-leaked data} Training using leaked data.}
\vspace*{-7mm}
\end{figure}

\vspace*{-5mm}
\section{Introduction}
\label{sec:intro}
\vspace*{-2mm}
% \joshua{Can I use the same citations from our previous papers here? Similar equations when summarizing prior work?} 
% \saurabh{Yes}
With growing concerns of data privacy, federated learning (FL)~\cite{mcmahan2017communication} has gained traction as a potential privacy-preserving method for training machine learning models. Compared to centralized learning where training is done on data localized at a central server, FL takes a decentralized approach where participating clients train a model on their local data and send their updates to the server. A typical training round involves a server sending a model to the clients, the clients training the model using their local data, and finally having the clients send their updates to the server for aggregation. However, the privacy-preserving property of FL only holds if the updates cannot be used to extract sensitive information about the local training data.

Despite only sending updates, prior works have shown the ability of attackers to gain information about the private training data through membership inference attacks~\cite{shokri2017membership,choquette2021label,nasr2019comprehensive}, property-inference attacks~\cite{melis2019exploiting,luo2021feature}, or GAN-based methods~\cite{hitaj2017deep,wang2019beyond}. Within these privacy attacks, data reconstruction has stood out as the most powerful, allowing attackers to directly recover the training data of clients~\cite{wen2022fishing,pasquini2021eluding}. Within data reconstruction attacks, gradient inversion (optimization-based attacks)~\cite{zhu2019deep,yin2021see,geiping2020inverting} and linear layer leakage~\cite{zhao2023loki, fowl2022robbing,boenisch2021curious} attacks stand out as the most common. 

While prior works have discussed the power and the limitations of these attacks in the context of an attacker simply breaching privacy~\cite{zhao2023resource,huang2021evaluating}, to the best of our knowledge, none has discussed the effectiveness of using the leaked data for downstream model training. Figure~\ref{fig:train-leaked data} shows this process of training using leaked data.
% \saurabh{The above is the punch line and comes way too late. We should bring this up (meaning this entire paragraph) and above the discussion that we now have earlier about what gradient inversion and LLL attacks are. That discussion should still be there but come later.}
% \joshua{I have moved it above the grad inversion and LLL discussions to right below where we first mention data reconstruction.}
% \saurabh{This is good.}
With the onset of deep learning, data has become a valuable commodity. This growth in AI also means that the value of data no longer lies just in breaching privacy of the raw data, but in the ability of the data to be used downstream for creating powerful models. Therefore, one measure of the success of a data reconstruction attack should be whether the leaked data is useful for machine learning training, an aspect omitted in prior work.
% \saurabh{The above is a key claim (which I wrote). Is this correct?}
% \joshua{Yes, this covers the goal of this paper. Is it too strong say the "value of data no longer lies in breaching privacy"?}
% \saurabh{Reworded a little.}

In order to discuss training on leaked data, we first introduce the attacks. Gradient inversion (GI) attacks typically reconstruct data through an iterative process where the distance between a gradient computed from a dummy image and the ground truth gradient is minimized by an optimizer. Label reconstruction, the process of recovering the batch labels from the client gradient, also plays a critical role in affecting the final reconstruction quality. 
% SB (11/17/23): This would be an apt place to define label reconstruction. 
% JZ (11/17/23): Added a simple definition at the beginning of the sentence.
These labels are typically leaked prior to optimization from the gradients of the final layer of the network~\cite{geng2021towards,yin2021see,zhao2020idlg,ma2022instance} and are matched with reconstructed images during the optimization. These attacks have shown success with individual updates of smaller batch sizes, up to 48 on ImageNet~\cite{yin2021see} or 100 on CIFAR~\cite{geiping2020inverting}. However, this only considers breaching privacy, as a batch size of 48 on ImageNet or 100 on CIFAR result in low quality reconstructions and/or a very small number of identifiable images. Even for a modest batch size of 8 or 16 on CIFAR-10, some batch reconstructions already fail. 
% SB (11/16/23): This sentence seems to contradict what we have above of being able to handle batch sizes of 100 on CIFAR. 
% JZ (11/16/23): Yes, I agree. I have added some clarification.
% SB (11/17/23): This is now good.
% In prior work, leakage has been quantified in the context of breaching privacy by whether there are any identifiable images in the entire reconstructed batch. 
Given the importance of data quality for training models, a sufficient condition for success is no longer if any images are identifiable, but rather the overall usefulness of the leaked data for the downstream training task. Here, even reconstructions with low similarity scores to the ground truth may still positively contribute to training.
% SB (11/16/23): The previous sentence is not clear. 
% JZ (11/16/23): Rephrased.

Compared to GI, linear layer leakage (LLL) does not suffer from problems with reconstruction quality. Using modification of a fully-connected (linear, FC) layer, an attacker can directly recover a proportion of the images of a client batch~\cite{zhao2023loki,fowl2022robbing,boenisch2021curious}. Reconstructions are done by solving a simple linear equation and have little computation overhead. However, these attacks typically require modification of the model architecture and can add a large overhead in terms of model size~\cite{zhao2023resource}. While data quality is not an issue, when using the leaked data for training models, labeling the data becomes a problem.
% \saurabh{The concept of label leakage may not be familiar to some reviewers and needs to be defined earlier when we first talk of that.}
% \joshua{I introduced it briefly in the previous part with gradient inversion. I added a clarification here that it means the same thing as in gradient inversion.}
% \saurabh{I added a comment where we talk of label leakage under GI.}
Unlike the optimization attacks, even if labels are known prior to reconstruction, the LLL process does not match them to the leaked images. As a result, after reconstruction there would be a set of leaked images and labels, where the images are not matched with their corresponding label.
% SB (11/16/23): You mean label for image 1 in batch A can erroneously get put on image 2 in the same batch A?
% JZ (11/16/23): I have reworded it. For clarification, what I mean is given a set of labels, the optimization process will match the reconstructions to corresponding labels during the actual optimization. As a result, leaked images will be matched with labels. However, linear layer leakage does not have this matching process. So even if we get the batch labels, the reconstruction process of linear layer leakage does not match them to the leaked images. So we will just have a set of leaked images and a set of labels and we won't know which ones go with which.
% SB (11/17/23): Agreed
Furthermore, because only a proportion of images in the given batch are leaked for LLL, even with a full set of labels, some labels will not have a leaked image to go with. This can lead to a tedious process of manually labeling leaked images. 
% SB (11/16/23): This last problem exists with gradient inversion attack also, and is in fact more severe as a smaller fraction of images is leaked. 
% JZ (11/16/23): Reworded a bit. For label leakage, if done correctly, the entire batch of labels will be leaked by the gradient. So even if gradient inversion has low quality reconstructions on that batch size, the labels will be matched with the images (albeit they may be very low quality reconstructions). However, for linear layer leakage, maybe only 85% of the batch will actually be leaked. Since we have all labels, 15% of them wouldn't correspond to any images (as the images were not leaked).

Viewing data reconstruction attacks through the lens of training models brings up many questions. How do models trained with leaked images compare to centralized or federated learning? Does the reconstruction quality impact how well models perform when trained on leaked data? How does the lack of label matching during reconstruction affect linear layer leakage? 
% Will the proportion of images leaked affect training? 
% SB (11/16/23): The above seems obvious - more images leaked means more data, more data means better training. How about a question about label leakage?
% JZ (11/16/23): Okay, replaced.
In this work we address these largely unanswered questions and highlight the problems that arise in this new setting. Our contributions go as follows:
\begin{itemize}
    \item We demonstrate the effectiveness of training models using leaked data compared to federated learning and the centralized baselines. Using the data from linear layer leakage and optimization, models trained on leaked data can achieve $20.40\%$ and $17.58\%$ higher accuracy on CIFAR-10 compared to federated learning.
    % SB (11/16/23): Compared to FL?
    % JZ (11/16/23): Yes.
    \item We show that similarity metrics such as PSNR do not show if reconstructed images are useful for training, as even some of the worst images improve models. Training on CIFAR-10 with bad reconstructions generated by Inverting Gradients (batch size 16) with PSNR $<14$ results in a $58.29\%$ test accuracy. Removing these images and training only on images $>14$ PSNR results in only a minor accuracy decrease from $76.83\%$ to $75.48\%$. 
    % SB (11/16/23): An absolute number like this is less useful. Compared to a PSNR of XXX, where the model accuracy is only a slight improvement, at XXX\%. 
    % JZ (11/16/23): Modified the contribution to be more clear.
    \item We quantify the effects and highlight the crucial limitations of current data reconstruction attacks when viewed through the lens of downstream model training. While gradient inversion can breach privacy for a batch size of 100 on CIFAR-10, this is impractical for training as even increasing the batch size from 4 to 16 results in a performance drop from $90.34\%$ to $76.83\%$. 
    % \joshua{Is this contribution okay? I removed the LLL part since it was long}
    % \saurabh{I like this.}% For linear layer leakage, when only a few labels are known, accuracy suffers. Going from 250 to 20 labels using semi-supervised learning results in an accuracy drop from $91.65\%$ to $86.69\%$.
    % SB (11/16/23): Above is not clear. We need to bring this point out earlier in the introduction, and here we need to quantify like we do for the other points. 
    % JZ (11/16/23): Quantified a bit. Removed the part talking about LLL as the contribution was very long.
\end{itemize}

%% file: sec/2_related.tex
\vspace*{-2mm}
\section{Related work}
\label{sec:related}
\vspace*{-2mm}
Gradient inversion (GI) attacks operate under an honest-but-curious threat model where a server only knows the update and model. The method involves feeding a dummy image into a model and computing the subsequent gradient. The dummy image starts out initialized as random noise, and an optimizer iteratively minimizes the distance between the computed gradient and the ground truth client gradient.
\vspace*{-2mm}
\begin{equation}\label{eq:optim}
    x^* = \arg \min_{x}||\nabla L(x, y, \theta)-\nabla W||_2
\vspace*{-2mm}
\end{equation}
The intuition behind these attacks is that a similar computed dummy gradient $\nabla L(x, y, \theta)$ and ground truth gradient $\nabla W$ will result from similar images. Thus, as the gradients become closer, the final reconstructed image $x^*$ should become close to the ground truth training image. The label $y$ is assumed to be known prior to optimization. For a batch reconstruction with multiple labels, the labels are matched with the reconstructed images during optimization (as the loss and gradient are computed with these image-label pairs). 
% Minimizing the distance between the ground truth and generated gradient from these pairs results in reconstructions matched with a correct label.
% SB (11/16/23): The above sentence is not clear. 
% JZ (11/16/23): Rephrased
Regularizers have also been introduced to assist the optimization process~\cite{geiping2020inverting,yin2021see}. However, these can result in reconstructed image artifacts.

Linear layer leakage (LLL) attacks are built on the observation that a fully-connected (FC) layer leaks the input to the layer through the gradients~\cite{phong2017privacy,fan2020rethinking}.
\vspace*{-2mm}
\begin{equation}\label{eq:1}
    x^i = \frac{\delta L}{\delta W^i} / \frac{\delta L}{\delta B^i}
\vspace*{-2mm}
\end{equation}
\noindent
Here, $\frac{\delta L}{\delta W^i}$ is the weight gradient and $\frac{\delta L}{\delta B^i}$ is the bias gradient of a neuron. Neuron $i$ is activated by an image and $x^i$ is the respective reconstructed image. However, this assumes that only a single image activated the neuron. If multiple images activate the same neuron, reconstructed image $x^i$ becomes a combination of the images.

To mitigate this, the methods of trap weights~\cite{boenisch2021curious} and Robbing the Fed~\cite{fowl2022robbing} were proposed. Trap weights initializes the weights of the FC layer randomly as half positive and half negative, where the negative parameters are sampled from a slightly larger magnitude. This method has difficulty with scaling~\cite{zhao2023loki} compared to Robbing the Fed, which proposed a method where the FC layer weights are used to measure a property of the images (e.g., average pixel brightness). The biases of the layer were then set to fit the distribution of the dataset and a ReLU activation was used to threshold the activation. Images were reconstructed as
\vspace*{-2mm}
\begin{equation}\label{eq:2}
    x^i = (\frac{\delta L}{\delta W^i} - \frac{\delta L}{\delta W^{i+1}}) / (\frac{\delta L}{\delta B^i} - \frac{\delta L}{\delta B^{i+1}})
\vspace*{-2mm}
\end{equation}
where $i+1$ indicates the neuron with a bias being the next highest cutoff. This method requires basic knowledge about the input data distribution, but achieves a higher leakage rate compared to trap weights and is scalable to aggregation. 

In the case of FedAvg, the sparse variant of Robbing the Fed utilizes a two-sided activation function (such as Hard tanh) and weight/bias scaling to maintain the same leakage as in FedSGD. However, the sparse variant is not scalable and leads to precision problems when attacking secure aggregation~\cite{bonawitz2017practical} or larger batches. The LOKI attack addresses the scalability problems of both methods through the introduction of a convolutional layer that splits the scaling between number of clients and batch size in secure aggregation~\cite{zhao2023loki}. They further introduce a convolutional scaling factor (CSF) which achieves a higher leakage rate in FedAvg without increasing the FC layer size. 

While there have been many works exploring data reconstruction, prior work has not evaluated these attacks in the context of downstream tasks. The works on GI~\cite{zhu2019deep,geiping2020inverting,yin2021see} and LLL~\cite{fowl2022robbing,boenisch2021curious,zhao2023loki} attacks only discuss the reconstruction quality of the methods in terms of standard image metrics such as PSNR (peak signal-to-noise ratio), SSIM (structural similarity index measure), or LPIPS (learned perceptual image patch similarity ). 
% GradInversion~\cite{yin2021see} also introduced the IIP (image indentifiability precision) metric. 
However, these metrics only measure how similar the reconstruction are to the ground truth and only give a vague sense of how useful the data is for training. Another work discussed the limitations of GI attacks~\cite{huang2021evaluating} measured by image similarity. In~\cite{zhao2023resource}, LLL attacks were measured in terms of the resource overhead added by the attacks. However, these works also did not discuss the usefulness of reconstructed data post-leakage.

This work aims to bridge the gap and explore the important facet of training machine learning models on leaked data from reconstruction attacks. Viewing leaked data beyond privacy breach brings to light key weaknesses of current attacks in generating useful data. While LLL reconstructs high quality images, the other important half of data, namely having matching labels, is missing. GI attacks can breach privacy for larger batch sizes, but the quality of the leaked data would be nearly useless for training models. In the age of machine learning and artificial intelligence, the value of data has increased tremendously. Therefore, when looking at leaked data, it is important to measure success in terms of usefulness to model training.

\saurabh{So the key question is has any prior work evaluated the function of leaked images on accuracy of downstream task? If not, we should make a bold statement to that effect at the end.}
\joshua{Yes, we have not found any prior work to explore this yet. Added the paragraph above.}

%% file: sec/3_methodology.tex
\vspace*{-2mm}
\section{Training on leaked data}
\label{sec:methodology}
\vspace*{-2mm}
\joshua{Some parts mentioned in related work/intro.}
Data reconstruction attacks have largely focused on the quality of the recovered image in gradient inversion (GI) or the efficiency of linear layer leakage (LLL). However, label reconstruction is important for training on leaked data. Within the pipeline of GI attacks, labels are typically recovered prior to reconstruction, as they assist in the optimization problem~\cite{huang2021evaluating}. % Prior methods such as Inverting Gradients~\cite{geiping2020inverting} and GradInversion~\cite{yin2021see} were originally limited in that label restoration was only possible without duplicate classes. 
% SB (11/16/23): I don't understand the above sentence. 
% JZ (11/16/23): Rephrased and moved after the next sentence. 
Recent works have shown that label restoration with large batches and duplicate labels is possible~\cite{geng2021towards,ma2022instance}. More accurate labels and the added flexibility of duplicate class images directly improves prior GI attacks such as Inverting Gradients~\cite{geiping2020inverting} and GradInversion~\cite{yin2021see}.

Interestingly, one strength of LLL has been in the ability to reconstruct images without knowledge of the data labels. While the class labels can affect the distribution of the client dataset (e.g., average image brightness can be different between classes)~\cite{zhao2023loki}, the labels themselves do not help in the reconstruction. Knowing labels helps guide GI attacks (as gradients are computed from data and label pairs), but for LLL, images activating the neurons are reconstructed using only the weight and bias gradients of the FC layer as with Equation~\ref{eq:2}. % Furthermore, labels are leaked from the client updates. Since the linear layer leakage attack would already be embedded into the update, leaking the labels afterward would not contribute to the reconstruction of the images. This is only the case for privacy breaching.
% SB (11/17/23): I do not follow the above sentence. 
% JZ (11/17/23): I have replaced the previous two sentences as they are confusing.
Labels are necessary to train models using leaked data, but even with a set of recovered labels, the leaked images would still remain unmatched. Unlike GI which matches during optimization, images leaked by LLL would need to be manually matched. For this work, we experiment with cases when all labels are known and matched and alternatively, when only a small portion of labels are known and matched. 
% SB (11/17/23): What about matching of labels, not just knowing the labels? For the above, do we mean: "For this work, we experiment with cases where all labels are known and matched and alternately, when only a portion of labels are known and matched."
% JZ (11/17/23): Yes, that is correct. Updated.
Nonetheless, the label leakage and matching issue stands as an open problem for LLL attacks.

Identifying which images are reconstructed correctly is also a challenging problem for LLL. For FedSGD, bins without activation can be identified, as after the subtraction process the reconstruction will be zero. However, if multiple images activate the same bin, the reconstruction is non-zero. By checking for non-zero values, we cannot directly identify whether the reconstruction is a single image or a combination of multiple. Manually filtering this can become a tedious process in practice. This is somewhat mitigated in the FedAvg case by LOKI. Across separate FedAvg local mini-batches, multiple images will {\em not} activate the same neuron, resulting in much fewew overlapping reconstructions. 
% SB (11/17/23): Changed "within" to "across". 
% JZ (11/17/23): Okay
Therefore the only images that can cause multiple activations are within the same mini-batch~\cite{zhao2023loki}. In this work, we manually filter out multiple image reconstructions in order to test the quality of the data. However, especially for LLL in FedSGD, the problem of how to automate removal or how to use these multiple image reconstructions for training still remains an important item of future work. 

%% file: sec/4_experiments.tex
\vspace*{-7mm}
\section{Experiments}
\label{sec:experiments}
\vspace*{-2mm}
The goal of our work is to discuss the effectiveness of leaked data when used for the downstream task of image classification. We use centralized training on the entire benign dataset and federated learning (with only updates and no data shared by the clients) as the baselines for comparison. Typically, these serve as the upper bound and the lower bound respectively for the model accuracy. For federated learning, we use two settings of 10 and 50 clients. We use a non-IID bias of 0.5 and FedAvg with 3 local iterations. Additional training results for the IID setting and FedSGD are included in the supplementary.

We use Inverting Gradients~\cite{geiping2020inverting} as the gradient inversion (GI) attack. Another candidate is GradInversion, which is a SOTA GI attack (especially for ImageNet-sized images). However, we discarded this as the author code for this attack is not available and the third-party online implementations did not achieve comparable results to the original paper. For linear layer leakage (LLL), we use the LOKI attack~\cite{zhao2023loki}. We use LOKI as it achieves SOTA FedAvg leakage rate and equivalent SOTA leakage rate in FedSGD compared to Robbing the Fed. Model performance for other LLL methods~\cite{fowl2022robbing,boenisch2021curious} can be found in the supplementary. 

We use MNIST~\cite{lecun1998mnist}, CIFAR-10~\cite{krizhevsky2009learning}, and the Tiny ImageNet~\cite{le2015tiny} datasets. For the MNIST dataset, we train using a simple DNN with 2 convolutional layers and 2 FC layers. For CIFAR-10 and Tiny Imagenet, we use a ResNet-18~\cite{he2016deep}. We evaluate on the validation set for Tiny ImageNet because the test set is unlabeled. The plots only show a single run for each setting, but the (numerical) testing performance is reported as an average over 5 runs. We run all experiments on NVIDIA A100 80GB GPUs. 

\vspace*{-2mm}
\subsection{Gradient inversion computation time}
\vspace*{-2mm}
\begin{table}[]
\scriptsize
\begin{center}
\begin{tabular}{|lccc|}
\hline
\multicolumn{4}{|c|}{MNIST}                                                                                                                                                                                                                       \\ \hline
\multicolumn{1}{|l|}{Batch size} & \multicolumn{1}{c|}{Model}                      & \multicolumn{1}{c|}{\begin{tabular}[c]{@{}c@{}}Time per\\ batch (s)\end{tabular}} & \begin{tabular}[c]{@{}c@{}}Time for entire\\ dataset (days)\end{tabular} \\ \hline
\multicolumn{1}{|l|}{8}          & \multicolumn{1}{c|}{\multirow{3}{*}{DNN}}   

& \multicolumn{1}{c|}{54.79}                                                        
&  4.76                                                                     \\

\multicolumn{1}{|l|}{16}         & \multicolumn{1}{c|}{}                           & \multicolumn{1}{c|}{62.34}                                                        & 2.71                                                                     \\
\multicolumn{1}{|l|}{32}         & \multicolumn{1}{c|}{}                           & \multicolumn{1}{c|}{64.32}                                                        & 1.40                                                                     \\ \hline
\multicolumn{4}{|c|}{CIFAR-10}                                                                                                                                                                                                                    \\ \hline
\multicolumn{1}{|l|}{Batch size} & \multicolumn{1}{c|}{Model}                      & \multicolumn{1}{c|}{\begin{tabular}[c]{@{}c@{}}Time per\\ batch (s)\end{tabular}} & \begin{tabular}[c]{@{}c@{}}Time for entire\\ dataset (days)\end{tabular} \\ \hline
\multicolumn{1}{|l|}{4}          & \multicolumn{1}{c|}{\multirow{3}{*}{ResNet-18}} & \multicolumn{1}{c|}{422.81}                                                       & 61.17                                                                    \\
\multicolumn{1}{|l|}{8}          & \multicolumn{1}{c|}{}                           & \multicolumn{1}{c|}{437.54}                                                       & 31.65                                                                    \\
\multicolumn{1}{|l|}{16}         & \multicolumn{1}{c|}{}                           & \multicolumn{1}{c|}{435.33}                                                            & 15.75                                                                        \\ \hline
\end{tabular}
\end{center}
\vspace*{-5mm}
\caption{\label{tab:ig-comp-time} Computation time for Inverting Gradients running on a NVIDIA A100 80GB GPU. Small batches have more total batches and take more time for the whole dataset. Computation time is significantly higher for CIFAR-10 given the more complex model.}
\vspace*{-5mm}
\end{table}

Compared to LLL attacks, the computational overhead of GI attacks when reconstructing data, while done offline, is still non-negligible. In order to fully evaluate the leaked data, we ran the Inverting Gradients atack on the entire dataset of MNIST with batch sizes of 8, 16, and 32. We also ran over CIFAR-10 with batch sizes of 4, 8, and 16. We run 10,000 iterations of optimization for each batch. We evaluate the Inverting Gradients attack in the best case scenario with an untrained model where {\em all} labels are correctly recovered prior to optimization. 

Table~\ref{tab:ig-comp-time} shows the amount of time taken by Inverting Gradients run on a NVIDIA A100 80GB GPU for MNIST and CIFAR-10 on a DNN and ResNet-18 respectively. There is some variance in the average amount of time based on batch size, but when attacking the entire dataset, smaller batch sizes always take longer time as the number of total batches is much larger. CIFAR-10 takes significantly longer to reconstruct each batch given the more complex model of ResNet-18 compared to a DNN. For a batch size of 4, CIFAR-10 takes 422.81 seconds per batch. Iterating across the entire dataset on a single GPU takes 61.17 days.
% With MNIST, the time required for reconstructing a batch size of 16 was roughly 62.34 seconds while a batch size of 32 took 64.32 seconds. Using a single A100 GPU, iterating across the dataset takes roughly 2.71 and 1.40 days respectively. 
% While each batch of size 32 took slightly longer on average than of size 16, the larger total number of batches in the dataset using batch size 16 meant the smaller batch size took longer in total.

% While MNIST took a reasonable amount of time for optimization, CIFAR-10 took significantly longer due to the higher dimensionality of the images and the additional model complexity of using a ResNet-18 compared to a DNN. With a batch size of 4, each batch took on average 422.81 seconds. For a batch size of 8 this was 437.54 seconds and $\color{red}{\boldsymbol{422.81}}$ seconds for batch size 16. For a batch size of 4, running optimization across the entire dataset takes roughly 61.17 days. For a batch size of 8 and 16, it would take 31.65 days and $\color{red}{\boldsymbol{18.12}}$ days respectively.
\saurabh{Much of this text information about times (per batch and total) can be better presented through a table. Keep the explanation in the text in that case.}
\joshua{Added a table and replaced the explanation in the text with a slightly shorter one covering the main points.}

\vspace*{-2mm}
\subsection{Leakage statistics}
\vspace*{-2mm}

\begin{table}
  \centering
  \begin{subtable}[c]{1.0\linewidth}
\scriptsize
\begin{center}
% \begin{tabular}{|lcc|}
% \hline
% \multicolumn{3}{|c|}{CIFAR-10}                                       \\ \hline
% \multicolumn{1}{|l|}{Batch size} & \multicolumn{1}{c|}{PSNR$\uparrow$}  & SSIM$\uparrow$ \\ \hline
% \multicolumn{1}{|l|}{4}          & \multicolumn{1}{c|}{27.88} & 0.9067 \\
% \multicolumn{1}{|l|}{8}          & \multicolumn{1}{c|}{20.77} & 0.7215 \\
% \multicolumn{1}{|l|}{16}         & \multicolumn{1}{c|}{15.94} & 0.4978 \\ \hline
% \multicolumn{3}{|c|}{MNIST}                                          \\ \hline
% % \multicolumn{1}{|l|}{Batch size} & \multicolumn{1}{c|}{PSNR$\uparrow$}  & SSIM$\uparrow$ \\ \hline
% \multicolumn{1}{|l|}{16}         & \multicolumn{1}{c|}{19.15} & 0.6506 \\
% \multicolumn{1}{|l|}{32}         & \multicolumn{1}{c|}{15.78} & 0.4537 \\ \hline
% \end{tabular}
\begin{tabular}{|lcc|ccc|}
\hline
\multicolumn{3}{|c|}{CIFAR-10}                    & \multicolumn{3}{c|}{MNIST}                          \\ \hline
\multicolumn{1}{|l|}{Batch size} & PSNR  & SSIM   & \multicolumn{1}{c|}{Batch size} & PSNR    & SSIM    \\ \hline
\multicolumn{1}{|l|}{4}          & 27.88 & 0.9067 & \multicolumn{1}{c|}{8}          
& 23.16 & 0.6996 \\
\multicolumn{1}{|l|}{8}          & 20.77 & 0.7215 & \multicolumn{1}{c|}{16}         & 19.15   & 0.6506  \\
\multicolumn{1}{|l|}{16}         & 15.94 & 0.3978 & \multicolumn{1}{c|}{32}         & 15.78   & 0.4537  \\ \hline
\end{tabular}
\end{center}
    \vspace*{-3mm}
    \caption{Gradient Inversion PSNR/SSIM}
    \vspace*{1mm}
    \label{tab:ig-psnr-ssim}
  \end{subtable}
  \hfill
    \begin{subtable}[c]{1.0\linewidth}
\scriptsize
\begin{center}
\begin{tabular}{|l|cc|cc|cc|}
\hline
                                                    & \multicolumn{2}{c|}{CIFAR-10}                                     & \multicolumn{2}{c|}{MNIST}                                        & \multicolumn{2}{c|}{Tiny ImageNet}                                \\ \hline
\begin{tabular}[c]{@{}l@{}}FC\\ factor\end{tabular} & \begin{tabular}[c]{@{}c@{}}Leaked\\ images\end{tabular} & Percent & \begin{tabular}[c]{@{}c@{}}Leaked\\ images\end{tabular} & Percent & \begin{tabular}[c]{@{}c@{}}Leaked\\ images\end{tabular} & Percent \\ \hline
8                                                   & 43788                                                   & 87.58   &    52548                                                     &    87.58     & 85804                                                   & 85.80   \\
4                                                   & 39464                                                   & 78.93   & 45966                                                   & 76.61   & 75149                                                   & 75.15   \\
2                                                   & 29882                                                   & 59.76   & 35795                                                   & 59.66   & 58012                                                   & 58.01   \\
1                                                   & 18242                                                   & 36.48   & 21815                                                   & 36.36   & 35393                                                   & 35.39   \\ \hline
\end{tabular}
\end{center}
    \vspace*{-3mm}
    \caption{Linear layer attack leakage rate}
    \label{tab:loki-leak-rate}
  \end{subtable}
  \vspace*{-3mm}
  \caption{\label{tab:leakage-stats} (a) Average PSNR$\uparrow$ / SSIM$\uparrow$ scores of gradient inversion (Inverting Gradients) reconstructions for various batch sizes and (b) number of leaked images/leakage rate of linear layer leakage attack (LOKI) for several FC layer sizes.}
% \caption{\label{tab:leakage-stats} (a) Average PSNR$\uparrow$ / SSIM$\uparrow$ scores of gradient inversion (Inverting Gradients) reconstructions for various batch sizes on CIFAR-10 and MNIST and (b) number of leaked images/leakage rate of linear layer leakage attack (LOKI) on CIFAR-10, MNIST, and Tiny ImageNet for several FC layer sizes.}
  \vspace*{-6mm}
\end{table}

For GI we use Inverting Gradients~\cite{geiping2020inverting}, and for LLL we use LOKI~\cite{zhao2023loki}. For Inverting Grads, we use a batch size of 4, 8, and 16 on CIFAR-10 and 8, 16 and 32 on MNIST. For LOKI, we choose the FC layer size based on the batch size. We vary this with factors of 1, 2, 4 and 8 on CIFAR-10, MNIST, and Tiny ImageNet. We use a client batch size of 64 for all datasets, so an FC size factor of 2 would mean $64\cdot2=128$ units in the FC layer. 

Table~\ref{tab:ig-psnr-ssim} gives the average PSNR and SSIM metrics (where a higher score indicates greater similarity with the ground truth) for the reconstructions across the entire dataset of CIFAR-10 and MNIST from Inverting Gradients. Increasing the batch size results in a lower PSNR and SSIM score for both CIFAR-10 and MNIST. 

Table~\ref{tab:loki-leak-rate} reports the number of leaked images for LOKI. For an FC factor of $4$, LOKI leaks $78.93\%$, $76.61\%$, and $75.15\%$ of images on the CIFAR-10, MNIST, and Tiny ImageNet datasets. Increasing the FC layer size increases the model size overhead, but results in a higher leakage rate as images are more likely to activate different neurons. Decreasing the size creates a smaller model size overhead but results in lower leakage rate. 

\saurabh{We cannot just point the reviewer to a result table. We have to interpret that for her. For PSNR/SSIM, give the calibration number of what it is for the original images. For LOKI result say that it is the highest among all attacks, how FC factor affects it and the size.}
\joshua{Added discussion. For the PSNR/SSIM calibration for original images, it is a bit strange since SSIM would be 1, but PSNR is $\infty$. I have left that out and just mentioned higher is better for now as other papers also do not mention it. Should we also mention that Robbing the Fed jointly achieves the highest leakage rate for FedSGD?}

\vspace*{-2mm}
\subsection{Training on leaked data from scratch}
\label{sec:experiments-train-from-scratch}
\vspace*{-2mm}

\begin{figure*}
  \centering
  \begin{subfigure}{0.33\linewidth}
    \includegraphics[width=0.99\columnwidth]{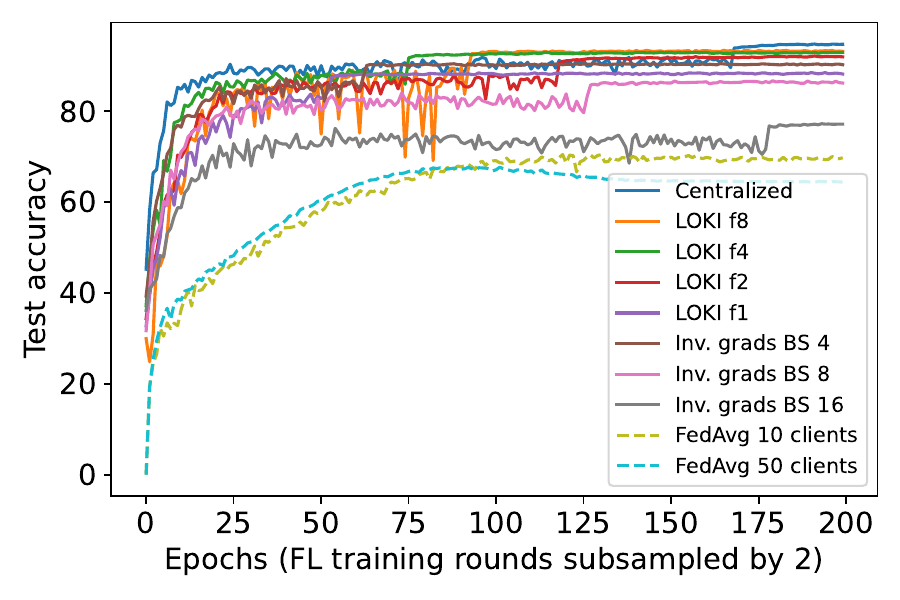}
    \caption{CIFAR-10}
    \label{fig:train-cifar}
  \end{subfigure}
  \hfill
    \begin{subfigure}{0.33\linewidth}
    \includegraphics[width=0.99\columnwidth]{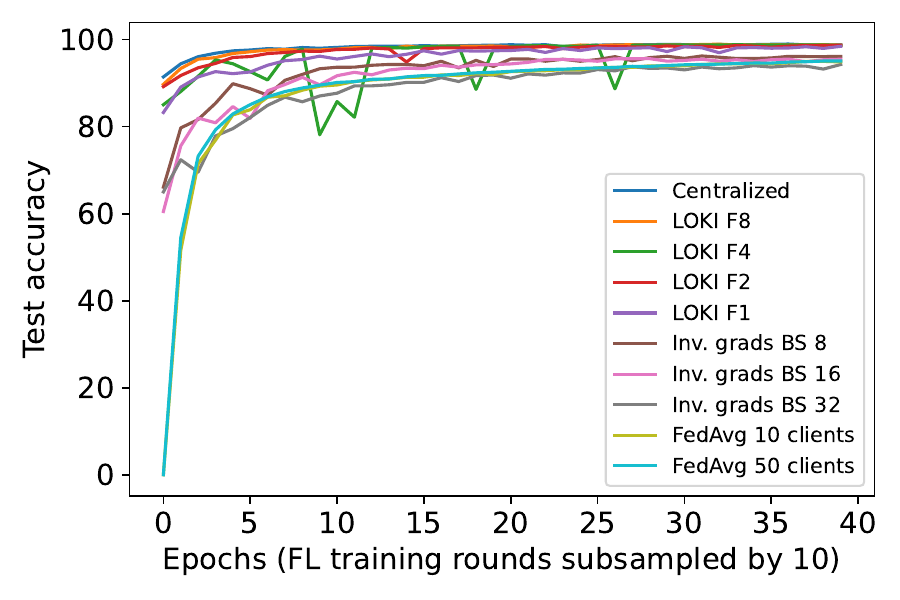}
    \caption{MNIST}
    \label{fig:train-mnist}
  \end{subfigure}
  \hfill
    \begin{subfigure}{0.33\linewidth}
    \includegraphics[width=0.99\columnwidth]{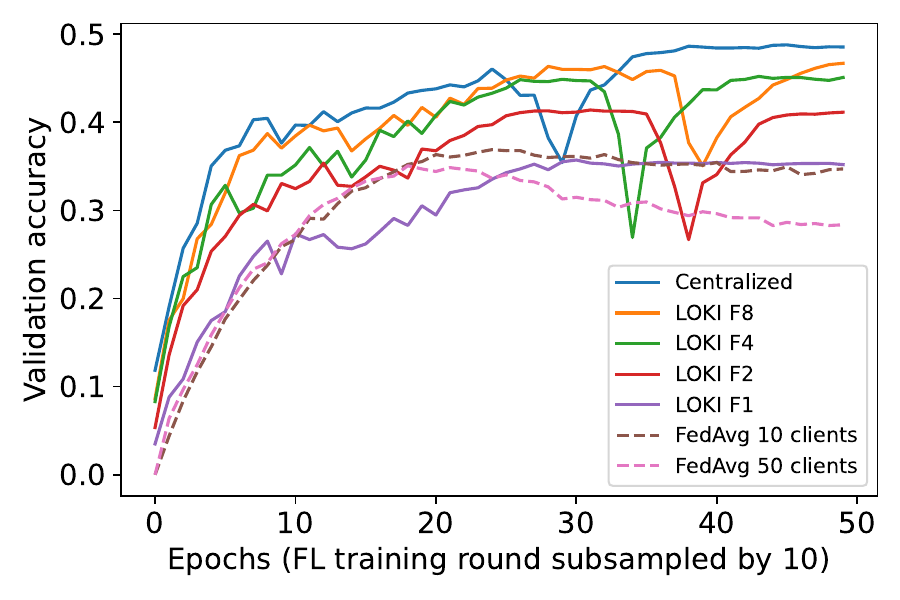}
    \caption{Tiny ImageNet}
    \label{fig:train-tinyim}
  \end{subfigure}
  \vspace*{-6mm}
  \caption{\label{fig:train-models} Training models on (a) CIFAR-10, (b) MNIST, and (c) Tiny ImageNet with leaked data compared to centralized and federated learning training. Both linear layer leakage and gradient inversion achieve higher accuracy than the federated learning (FedAvg) baseline for CIFAR-10 in all cases. For MNIST, LLL (LOKI) nearly reaches centralized accuracy while GI performs slightly worse than FL. Top-1 validation accuracy used when training models on Tiny ImageNet. Here, LLL performs better with a FC layer size factor of 2 or higher.}
  \vspace*{-3mm}
\end{figure*}

% \begin{figure}[!t]
% \begin{center}
% \includegraphics[width=0.9\columnwidth,trim={0 0 0 10mm},clip]{images/loki_train_cifar_factors.pdf}
% \end{center}
% \vspace*{-5mm}
% \caption{\label{fig:train-cifar} Training models on CIFAR-10 with leaked data compared to centralized and federated learning training. Using leaked data achieves higher accuracy than the FL baseline in all cases with gradient inversion and linear layer leakage.}
% \vspace*{-5mm}
% \end{figure}

% \begin{figure}[!t]
% \begin{center}
% \includegraphics[width=0.9\columnwidth,trim={0 0 0 10mm},clip]{images/FL_mnist.pdf}
% \end{center}
% \vspace*{-5mm}
% \caption{\label{fig:train-mnist} Training models on MNIST with leaked data compared to centralized and federated learning training. Linear layer leakage (LOKI) nearly reaches centralized accuracy while gradient inversion performs slightly worse than federated learning (FedAVG).}
% \vspace*{-5mm}
% \end{figure}

% \begin{figure}[!t]
% \begin{center}
% \includegraphics[width=0.9\columnwidth,trim={0 0 0 10mm},clip]{images/FL_tinyim_noseed_top1.pdf}
% \end{center}
% \vspace*{-5mm}
% \caption{\label{fig:train-tinyim} Top-1 validation accuracy when training models on Tiny ImageNet with leaked data from linear layer leakage compared to centralized and federated learning training. Leaked data performs better with a FC layer size factor of 2 or higher.}
% \vspace*{-5mm}
% \end{figure}

\begin{table*}
  \centering
  \begin{subtable}[c]{0.33\linewidth}
    \scriptsize
    \begin{center}
    \begin{tabular}{|l|c|c|}
    \hline
                                                                                                        & \begin{tabular}[c]{@{}c@{}}Num. clients /\\ Batch size /\\ FC factor\end{tabular} & \begin{tabular}[c]{@{}c@{}}Test \\ accuracy\end{tabular} \\ \hline
    Centralized                                                                                         & -                                                                                 & 94.38                                                   \\ \hline
    \multirow{2}{*}{FedAvg}                                                                             & 10                                                                                & 72.76                                                    \\
                                                                                                        & 50                                                                                & 68.71                                                    \\ \hline
    \multirow{3}{*}{\begin{tabular}[c]{@{}l@{}}Gradient inversion\\ (Inverting grads~\cite{geiping2020inverting})\end{tabular}} & 4                                                                                 & 90.34                                                    \\
                                                                                                        & 8                                                                                 & 86.16                                                    \\
                                                                                                        & 16                                                                                & 76.83                                                    \\ \hline
    \multirow{4}{*}{\begin{tabular}[c]{@{}l@{}}Linear layer leakage\\ (LOKI~\cite{zhao2023loki})\end{tabular}}              & {\bf 8}                                                                                 & {\bf 93.16}                                                   \\
                                                                                                        & 4                                                                                 & 92.94                                                  \\
                                                                                                        & 2                                                                                 & 91.90                                                   \\
                                                                                                        & 1                                                                                 & 88.86                                                   \\ \hline
    \end{tabular}
    \end{center}
    \vspace*{-3mm}
    \caption{CIFAR-10}
    \label{tab:train-cifar}
  \end{subtable}
  \hfill
    \begin{subtable}[c]{0.33\linewidth}
    \scriptsize
    \begin{center}
    \begin{tabular}{|l|c|c|}
    \hline
                                                                                                    & \begin{tabular}[c]{@{}c@{}}Num. clients /\\ Batch size /\\ FC factor\end{tabular} & \begin{tabular}[c]{@{}c@{}}Test \\ accuracy\end{tabular} \\ \hline
    Centralized                                                                                     & -                                                                                 & 98.89                                                    \\ \hline
    \multirow{2}{*}{FedAvg}                                                                         & 10                                                                                & 96.17                                                    \\
                                                                                                    & 50                                                                                & 96.18                                                    \\ \hline
    \multirow{3}{*}{\begin{tabular}[c]{@{}l@{}}Gradient inversion\\ (Inverting grads~\cite{geiping2020inverting})\end{tabular}} & 8    &   95.96              \\        
    %&        \textcolor{pink}{9}\textcolor{red}{9}\textcolor{orange}{9}\textcolor{yellow}{9}\textcolor{green}{9}\textcolor{cyan}{9}\textcolor{blue}{9}\textcolor{violet}{9}\textcolor{magenta}{9}                                             \\
    & 16                                                                                & 95.50                                                    \\
                                                                                                    & 32                                                                                & 94.29                                                    \\ \hline
    \multirow{4}{*}{\begin{tabular}[c]{@{}l@{}}Linear layer leakage\\ (LOKI~\cite{zhao2023loki})\end{tabular}}                                                                       
                                                                                                        & {\bf 8}                                                                                 &   {\bf 98.82}                                                   \\
                                                                                                        & 4                                                                                 & 98.70                                                    \\
                                                                                                    & 2                                                                                 & 98.72                                                    \\

                                                                                                    & 1                                                                                 & 98.46                                                    \\ \hline
    \end{tabular}
    \end{center}
    \vspace*{-3mm}
    \caption{MNIST}
    \label{tab:train-mnist}
  \end{subtable}
  \hfill
    \begin{subtable}[c]{0.33\linewidth}
    \scriptsize
\begin{center}
\begin{tabular}{|l|c|c|}
\hline
                                                                                        & \begin{tabular}[c]{@{}c@{}}Num. clients /\\ FC factor\end{tabular} & \begin{tabular}[c]{@{}c@{}}Validation \\ accuracy\end{tabular} \\ \hline
Centralized                                                                             & -                                                                  & 48.55                                                    \\ \hline
\multirow{2}{*}{FedAvg}                                                                 & 10                                                                 & 37.00                                                    \\
                                                                                        & 50                                                                 & 35.06                                                    \\ \hline
\multirow{4}{*}{\begin{tabular}[c]{@{}l@{}}Linear layer leakage\\ (LOKI~\cite{zhao2023loki})\end{tabular}} & \textbf{8}                                                                  & \textbf{46.70}                                                   \\
                                                                                        & 4                                                                  & 45.09                                                    \\
                                                                                        & 2                                                                  & 41.15                                                    \\
                                                                                        & 1                                                                  & 35.20                                                    \\ \hline
\end{tabular}
\end{center}
    \vspace*{-3mm}
    \caption{Tiny ImageNet}
    \label{tab:train-tinyim}
  \end{subtable}
  \vspace*{-4mm}
  \caption{\label{tab:train-models} (a) CIFAR-10 and (b) MNIST test accuracy, (c) Tiny ImageNet top-1 validation accuracy trained from scratch with leaked data. CIFAR-10 and Tiny ImageNet are trained with a ResNet-18 and MNIST is trained with a DNN. For the two attacks (gradient inversion and linear layer leakage), training happens with leaked data. Second column indicates the number of clients in federated learning (FedAvg), the batch size for the gradient inversion attack, and the fully-connected layer size factor for linear layer leakage. Best accuracy is used for FedAvg and final accuracy for all other settings.}
  \vspace*{-4mm}
\end{table*}

We start with a discussion on the models that can be trained from the leaked data of both GI and LLL attacks. The attacks are applied under the FedSGD setting. We also start with assuming that {\em all} labels are known prior to training --- this is a best case assumption for the downstream training. For CIFAR-10, we train the centralized models using SGD with momentum 0.9 and weight decay 0.0001. We use a batch size of 128 and train for 200 epochs starting with a initial learning rate of 0.1 and a scheduler that decreases on plateau. For MNIST, we use SGD without momentum or weight decay and train for 40 epochs with a LR of 0.01. For Tiny ImageNet we use Adam with learning rate 0.001 and train for 50 epochs with a batch size of 1024. For federated learning, we use 400 training rounds for CIFAR-10 and MNIST and 500 rounds for Tiny ImageNet. Client batch size of 32, 128, and 256 are used for MNIST, CIFAR-10, and Tiny ImageNet. The test accuracy is downsampled to fit on the same axis with centralized training.

Table~\ref{tab:train-cifar} and Figure~\ref{fig:train-cifar} show the results for CIFAR-10 for centralized training, federated learning (FedAvg), and training with leaked data. Centralized training achieves a $94.38\%$ accuracy and federated learning achieves a $72.76\%$ and $68.71\%$ peak accuracy with 10 and 50 clients respectively. Centralized and FL indeed provide the upper bound and the lower bound of the accuracy here. Across all attack settings, GI and LLL achieve higher model accuracy compared to federated learning. The reconstruction quality of GI is adversely affected by larger batch sizes and in turn also impacts the final model performance. With a batch size of 4, 8, and 16, models trained on the leaked data from GI achieves $90.34\%$, $86.16\%$, and $76.83\%$ accuracy. With even larger batch sizes, it is likely that the model performance will drop below federated learning. The performance of models using LLL data is more stable. Between an FC layer size of 512 (factor 8) and 64 (factor 1), the accuracy only drops from $93.16\%$ to $88.86\%$. This is spite of a large drop in the number of leaked images from 43788 ($87.58\%$) to 18242 ($36.48\%$) as shown in Table~\ref{tab:loki-leak-rate}. This is a subtle and consequential result --- the downstream training task can still be accomplished despite a big drop in the amount of data leakage. All prior work had stopped at the stage of evaluating the proportion of leakage and therefore had missed out on this insight. 
% SB (11/17/23): Verify above. 
% JZ (11/17/23): Yes, this is true.

Table~\ref{tab:train-mnist} and Figure~\ref{fig:train-mnist} show the results for MNIST. LLL performs better than the FedAvg baseline in all cases. With an FC size factor of 8, LOKI achieves a $98.82\%$ test accuracy (nearly the same regardless of the FC size factor), only $0.07\%$ lower than centralized at $98.89\%$. GI performs slightly {\em worse} than federated learning, achieving $95.96\%$, $95.50\%$, and $94.29\%$ at batch sizes 8, 16, and 32 compared to $96.17\%$ and $96.18\%$ with 10 and 50 clients in FedAvg. We believe this is because the noise in the reconstructions hampers model performance with GI. This affects MNIST more severely than for CIFAR because the MNIST images are more sparse and sensitive to noise. 
% even though images reconstructed for both batch sizes are still easily visually identifiable.

The top-1 validation accuracy for Tiny ImageNet are given in Table~\ref{tab:train-tinyim} and Figure~\ref{fig:train-tinyim}. The total number of leaked images hurts LLL more here than in the other two datasets. Between an FC layer size of 512 (factor 8) and 64 (factor 1), the validation accuracy drops from $46.70\%$ to $35.20\%$, an $11.5\%$ drop in performance. Federated learning achieves $37.00\%$ and $35.06\%$ accuracy with 10 and 50 clients respectively. FL with 10 clients has slightly better accuracy than LLL with an FC size factor of 1. However, any LLL setting with a larger FC layer achieves higher accuracy than both settings in federated learning.

\vspace*{-2mm}
\subsection{FedAvg with leaked images from LOKI}
\vspace*{-2mm}
\begin{table}[]
\scriptsize
\begin{center}
\begin{tabular}{|l|c|c|}
\hline
\begin{tabular}[c]{@{}l@{}}FC size\\ factor\end{tabular} & \begin{tabular}[c]{@{}c@{}}LOKI\\ FedSGD\end{tabular} & \begin{tabular}[c]{@{}c@{}}LOKI\\ FedAvg\end{tabular} \\ \hline
8                                                        & 87.58 (43788)                                         & {\bf 92.19 (46097)}                                         \\
4                                                        & 78.93 (39464)                                         & {\bf 85.98 (42989)}                                         \\
2                                                        & 59.76 (29882)                                         & {\bf 75.29 (37645)}                                         \\
1                                                        & 36.48 (18242)                                         & {\bf 58.39 (29196)}                                         \\ \hline
\end{tabular}
\end{center}
\vspace*{-5mm}
\caption{\label{tab:fedavg-fedsgd-leak-rate} Leakage rate $\%$ (leaked images) of LOKI in FedSGD and FedAvg on CIFAR-10 for several FC layer sizes. FedAvg with half the FC layer size has comparable leakage rate to FedSGD.}
\vspace*{-3mm}
\end{table}

\begin{table}[]
\scriptsize
\begin{center}
\begin{tabular}{|l|c|c|c|}
\hline
                                                                                        & \begin{tabular}[c]{@{}c@{}}FC factor\end{tabular} & \begin{tabular}[c]{@{}c@{}}FedSGD\\ accuracy\end{tabular} & \begin{tabular}[c]{@{}c@{}}FedAvg\\ accuracy\end{tabular} \\\hline
\multirow{4}{*}{\begin{tabular}[c]{@{}l@{}}Linear layer\\ leakage (LOKI~\cite{zhao2023loki})\end{tabular}} & 8                                                                  & 93.16      & \textbf{93.31}                                              \\
                                                                                        & 4                                                                  & \textbf{92.94}     & 92.88                                                \\
                                                                                        & 2                                                                  & 91.90      & \textbf{92.35}                                             \\
                                                                                        & 1                                                                  & 88.86      & \textbf{91.11}                                                \\ \hline
\end{tabular}
\end{center}
\vspace*{-5mm}
\caption{\label{tab:train-fedavg} CIFAR-10 test accuracy with LOKI FedAvg.}
% SB (11/17/23): Copy over the test accuracy column with LOKI FedSGD and reword caption. 
% JZ (11/17/23): Added
\vspace*{-5mm}
\end{table}

We explore the impact of leaking images using LLL in FedAvg when training models. We use 8 local iterations of mini-batch size 8, a local dataset size of 64, and a learning rate of 1e-4. Following the LOKI attack~\cite{zhao2023loki}, CSF$=500$ is used to achieve a higher leakage rate and reconstruction quality. Similar to before, we apply the attack across batches covering the entire CIFAR-10 training dataset. 

Table~\ref{tab:fedavg-fedsgd-leak-rate} shows the leakage rate of LOKI in FedAvg compared to FedSGD. Using $\textit{CSF}=500$, the leakage rate of LOKI in FedAvg is substantially higher than the leakage rates in the FedSGD setting. Using an FC layer of half the size, the leakage rate in FedAvg is comparable to that of FedSGD. For example, LOKI FedAvg factor 1 achieves a leakage rate of $58.39\%$ compared to LOKI FedSGD factor 2 which achieves a leakage rate of $59.76\%$. 

Table~\ref{tab:train-fedavg} shows the testing accuracy of LOKI FedAvg for the different FC layer sizes. The test accuracies are very comparable to the FedSGD settings. However, with a smaller FC layer size factor, the effect of having a larger total number of leaked images becomes visible. With FC factor 1, LOKI FedAvg achieves $91.11\%$ accuracy while LOKI FedSGD achieves $88.86\%$.

\vspace*{-2mm}
\subsection{Training with semi-supervised learning}
\vspace*{-2mm}

% \begin{figure}[!t]
% \vspace*{-1mm}
% \begin{center}
% \includegraphics[width=0.78\columnwidth,trim={0 0 0 10mm},clip]{images/40_labels_cifar10_top1.pdf}
% \end{center}
% \vspace*{-6mm}
% \caption{\label{fig:ssl-40} Semi-supervised learning using CoMatch on CIFAR-10 with a WideResNet. The number of leaked images decreases with smaller FC layer sizes and the number of known labels is fixed at 40.  A performance drop is observed with LOKI FC factor 1.}
% \vspace*{-7mm}
% \end{figure}

% \begin{figure}[!t]
% \begin{center}
% \includegraphics[width=0.78\columnwidth,trim={0 0 0 12mm},clip]{images/vary_labels_cifar10_top1.pdf}
% \end{center}
% \vspace*{-7mm}
% \caption{\label{fig:ssl-vary} Semi-supervised learning using CoMatch on CIFAR-10 with a WideResNet. Using LOKI with a FC size factor of 4, the number of known labels varies between 20, 40 and 250.}
% \vspace*{-4mm}
% \end{figure}

\begin{figure}
\vspace*{-1mm}
  \centering
  \begin{subfigure}{0.75\linewidth}
    \includegraphics[width=1\columnwidth]{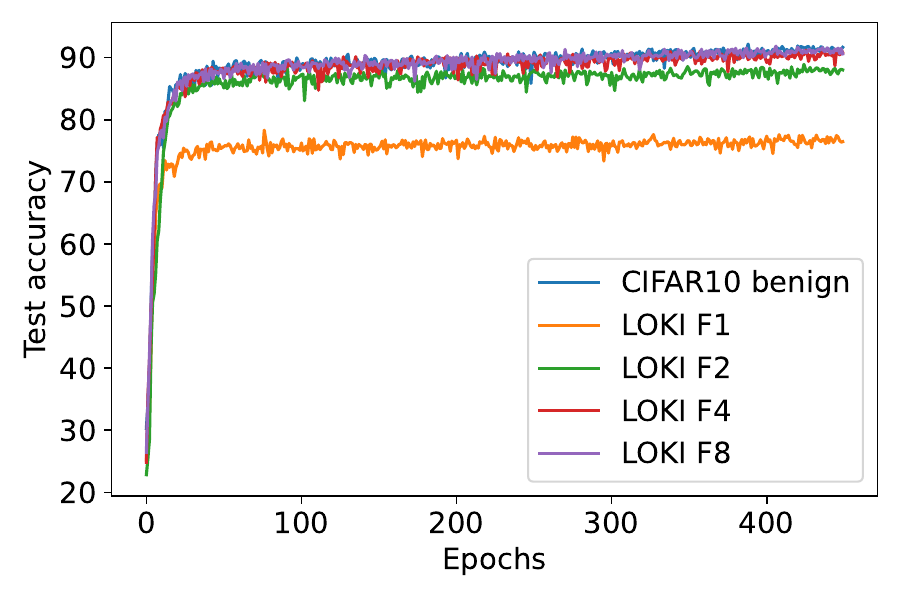}
    \caption{Varying leakage rate}
    \label{fig:ssl-40}
  \end{subfigure}
  \vspace*{3mm}
  % \hfill
    \begin{subfigure}{0.75\linewidth}
    \includegraphics[width=1\columnwidth]{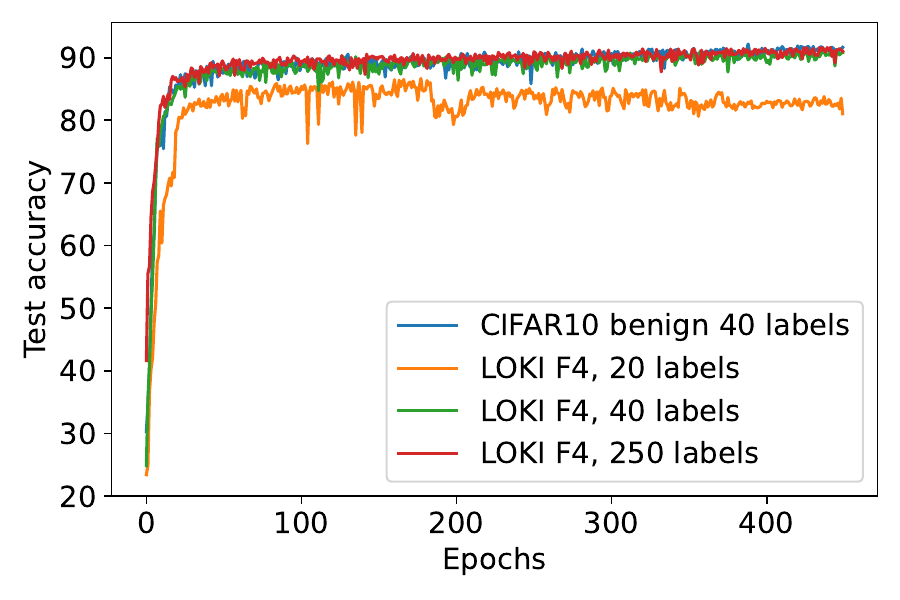}
    \caption{Varying number of  labels}
    \label{fig:ssl-vary}
  \end{subfigure}
  \vspace*{-5mm}
  \caption{\label{fig:ssl-all}Semi-supervised learning using CoMatch on CIFAR-10 with a WideResNet with (a) a varying FC size and leakage rate (LR) for LOKI and the known labels fixed at 40 and (b) a fixed number of leaked images and 20, 40, and 250 known labels.}
  \vspace*{-5mm}
\end{figure}

In order to train machine learning models in a supervised fashion, images leaked through LLL need to be labeled. Previously, we assumed that all labels were known and matched to the images prior to training, but this is impractical. Within a LLL batch reconstruction, the order of images is not preserved. Even if labels were recovered like for GI attacks, this would result in a set of unmatched ground truth labels and images. Manual hand labeling is tedious and this inability to match labels is a weakness of LLL attacks \emph{only when it comes to downstream tasks.}

In this experiment, we assume that only a small portion of images are labeled. This setting is similar to semi-supervised learning (SSL)~\cite{yang2022class,zheng2022simmatch} and we use CoMatch~\cite{li2021comatch}, a semi-supervised algorithm, to train the models. We use a WideResNet-28-2~\cite{zagoruyko2016wide} and train for 450 epochs on CIFAR-10. Figure~\ref{fig:ssl-40} fixes the number of known labels to be 40 and shows the test accuracy curve with different fully connected layer sizes. We note that when there is only a small proportion of leaked data, this leads to a pronounced decrease in performance in SSL. LOKI with an FC factor of 1 (with $36.48\%$ leakage rate) illustrates this, as it only achieves $78.31\%$ max accuracy compared to the CIFAR-10 40 label baseline of $92.18\%$. With a factor of 2, the accuracy improves and jumps to $88.91\%$ as the leakage rate also substantially increases to $59.76\%$. 
% SB (11/17/23): Because the leakage rate increases substantially to XXX\%.
% JZ (11/17/23): Added.
With factors 4 and 8 ($78.93\%$ and $87.58\%$ leakage rate), the accuracy is very close to the baseline at $91.18\%$ and $91.84\%$ respectively. Figure~\ref{fig:ssl-vary} shows the test accuracy plot of LOKI size factor 4 with the number of known labels being 20, 40, and 250. With 20 and 250 labels, LOKI factor 4 achieves $86.69\%$ and $91.65\%$ peak accuracy. We note that there is a greater variability in accuracy with a smaller number of labels, a trend consistent with other SSL methods~\cite{li2021comatch,zheng2022simmatch,yang2022class}.
% JZ (11/17/23): This trend is also consistent with prior work and we can see it in the results of the other SSL papers.
% SB (11/17/23): Say that then with cites. Showing our results match up with prior results adds credibility to our paper. 
% JZ (11/17/23): Added.

% \begin{figure}
%   \centering
%   \begin{subfigure}{0.49\linewidth}
%     \includegraphics[width=1.0\columnwidth,trim={5mm 0 5mm 10mm},clip]{images/40_labels_cifar10_top1.pdf}
%     \caption{SSL vary FC size}
%     \label{fig:ssl-40}
%   \end{subfigure}
%   \hfill
%     \begin{subfigure}{0.49\linewidth}
%     \includegraphics[width=1.0\columnwidth,trim={5mm 0 5mm 10mm},clip]{images/vary_labels_cifar10_top1.pdf}
%     \caption{SSL vary labels}
%     \label{fig:ssl-vary}
%   \end{subfigure}
%   \vspace*{-3mm}
%   \caption{\label{tab:ssl-all} Semi-supervised learning using CoMatch on CIFAR-10 with a WideResNet-28-2. In (a), the FC layer size factor of LOKI is varied. For (b), the FC layer size facor is fixed at 4 and the number of known labels varies between 20, 40, and 250.}
%   \vspace*{-3mm}
% \end{figure}

\vspace*{-2mm}
\subsection{Starting training from federated models}
\vspace*{-2mm}
\begin{figure}[!t]
\begin{center}
\includegraphics[width=0.75\columnwidth]{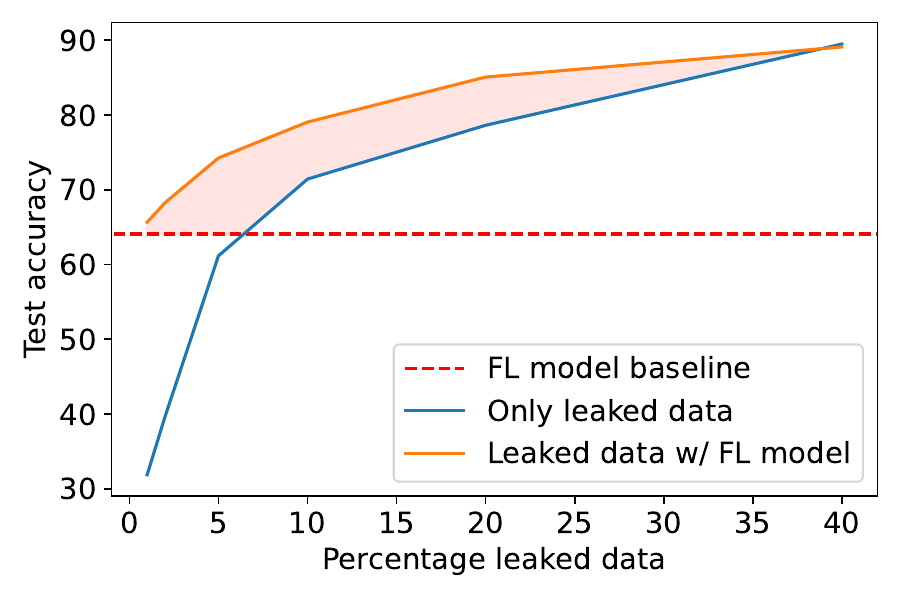}
\end{center}
\vspace*{-6mm}
\caption{\label{fig:train-w-fl-model} FL model trained with 50 clients as start point for training with leaked LLL data. Highlighted area indicates improvement above the FL model and training on the leaked data alone.}
\vspace*{-6mm}
\end{figure}

While the previous experiments worked with leaked data from all batches in the dataset, such large amounts of leaked data may not always be available. For example, LLL attacks may affect model performance due to manipulation of the model parameters and/or architecture. In later rounds when model performance is high, this may be easier to detect. Similarly, GI attacks such as Inverting Gradients~\cite{geiping2020inverting} have better reconstruction quality with an untrained network. In both cases, it is possible that the attack is only applied during a few initial rounds of training and only a small portion of data is leaked. Training from scratch with a small portion of leaked data does not achieve good model performance alone, so instead, we can initialize from the fully trained FL model and then train it centrally using the leaked data.
% SB (11/17/23): The punch line is missing here. "So instead of training from scratch using leaked data, we use the model that has been trained using FL and then fine tune using the leaked data."
% JZ (11/17/23): Added. I removed fine tuning, since it may not be the correct terminology.
% We illustrate this case using smaller proportions of the data leaked by LOKI.

We use smaller portions of the leaked data from LOKI FC size factor of 8 for this experiment. With only $5\%$ (2500) of leaked images in CIFAR-10, a model trained from scratch can only achieve $61.15\%$ test accuracy. With $2\%$ (1000 images) of data, the model performance drops to $39.63\%$. Both cases are below an FL model with 50 clients that achieves a $64.08\%$ final test accuracy. However, by using the FL model as the initial starting point, even a small amount of leaked data allows for an increase in performance. We take the model that has been trained using FL and then additionally train that model centrally. Using only $2\%$ and $5\%$ of the leaked data,
% SB (11/17/23): The above wording is confusing. What we mean is: We take the model that has been trained using FL and then fine tune that model centrally --- in the experiments we use respectively 2% and 5% of the leaked data. 
% JZ (11/17/23): Updated
the models are able to achieve $68.25\%$ and $74.25\%$ final testing accuracy, a $28.62\%$ and $13.10\%$ improvement from starting from scratch. Figure~\ref{fig:train-w-fl-model} shows the improvement in test accuracy achieved by using the trained FL model as initialization instead of starting from scratch. The improvement is especially noticeable with small percentages of leaked data.

\vspace*{-2mm}
\subsection{Quality of data reconstruction}
\label{sec:psnr-quality}
\vspace*{-2mm}

\begin{table}
  \centering
  \begin{subtable}{0.49\linewidth}
    \scriptsize
    \begin{center}
    \begin{tabular}{|l|c|c|}
    \hline
    PSNR             & \begin{tabular}[c]{@{}c@{}}\% imgs \\ kept\end{tabular} & \begin{tabular}[c]{@{}c@{}}Test \\ accuracy\end{tabular} \\ \hline
    $>20$ & 15.88                                                     & 70.04                                                    \\
    $>18$ & 27.59                                                     & 71.52                                                    \\
    $>16$ & 40.78                                                     & 73.73                                                    \\
    $>14$ & 60.22                                                     & 75.48                                                    \\
    $>12$ & 84.42                                                     & 76.16                                                    \\ \hline
    \end{tabular}
    \end{center}
    \vspace*{-1mm}
    \caption{PSNR above threshold}
    \label{tab:cifar-best-psnr}
  \end{subtable}
  \hfill
    \begin{subtable}{0.49\linewidth}
    \scriptsize
    \begin{center}
    \begin{tabular}{|l|c|c|}
    \hline
    PSNR             & \begin{tabular}[c]{@{}c@{}}\% imgs\\ kept\end{tabular} & \begin{tabular}[c]{@{}c@{}}Test \\ accuracy\end{tabular} \\ \hline
    $<20$ & 84.12                                                     & 73.03                                                    \\
    $<18$ & 72.41                                                     & 67.09                                                    \\
    $<16$ & 59.22                                                     & 63.71                                                    \\
    $<14$ & 39.78                                                     & 58.29                                                    \\
    $<12$ & 15.58                                                     & 45.05                                                    \\ \hline
    \end{tabular}
    \end{center}
    \vspace*{-1mm}
    \caption{PSNR below threshold}
    \label{tab:cifar-worst-psnr}
  \end{subtable}
  \vspace*{-3mm}
  \caption{\label{tab:train-models-psnr} Training models using CIFAR-10 leaked data from inverting gradients batch size 16. Only the reconstructions with a PSNR (a) above and (b) below the threshold are used in training.}
  \vspace*{-5mm}
\end{table}

Quality of reconstruction is an important concern for GI attacks. As the batch size continues to increase, the reconstruction quality decreases. Table~\ref{tab:ig-psnr-ssim} also shows this relationship. From the previous experiments, we also see a negative relationship between the increasing batch size and the usefulness of the leaked data in downstream model training. 
% Similarly, this shows that a relationship between the quality and usefulness of the data exists.
% SB (11/17/23): I chopped above. Verify.
% JZ (11/17/23): Correct

Here, we explore whether poor quality reconstructions are still useful for training models. Using the CIFAR-10 leaked dataset created using Inverting Gradients on a batch size of 16, we first sort the reconstructions by their PSNR value. Our first set of experiments removes the worst images from training. Table~\ref{tab:cifar-best-psnr} shows the percent of images above the PSNR threshold and the final test accuracy when trained on that data. Compared to the baseline accuracy of $76.83\%$ achieved by using the entire leaked dataset, even removing the worst images with a PSNR $<12$ from the training set results in a small but non-zero performance loss. 
Then we take a different look at this question by now training on the worst images. 
Table~\ref{tab:cifar-worst-psnr} shows the test accuracy of models trained on all images {\em below} a PSNR threshold. Even when training on the lowest quality images with a PSNR $<12$, the model achieves a $45.05\%$ final testing accuracy, significantly worse than the baseline, but much above zero.  
% SB (11/17/23): This is significantly worse accuracy than using all the leaked data. 

From the previous results, we see that even leaked data with a low reconstruction quality is still useful to the model training process. 
% While higher quality reconstructions are the ultimate goal, even the low quality images are useful towards training. 
This message is a new insight --- all prior works when calculating the leak rate do not count poorly reconstructed images. 
We additionally include results using SSIM in the supplementary section. Visual examples of poor quality reconstructions used to train the models are also included in the supplement.

\vspace*{-2mm}
\subsection{Observations on reconstruction quality trends}
\vspace*{-2mm}
\begin{figure}[!t]
\begin{center}
\includegraphics[width=0.9\columnwidth,trim={0 10mm 0 10mm},clip]{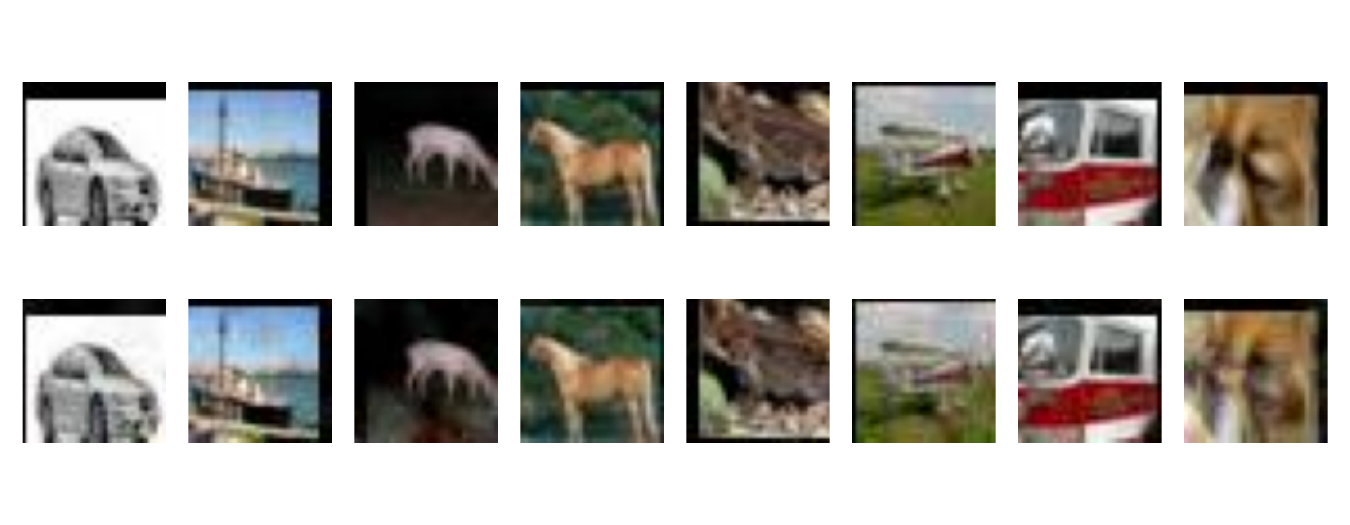}
\end{center}
\vspace*{-5mm}
\caption{\label{fig:ig-good-example} Reconstructions on CIFAR-10 from Inverting Gradients batch size 8. Ground truth images are on top and reconstructions are on the bottom. All labels in the batch are different and reconstructed images are high quality.}
\vspace*{-5mm}
\end{figure}

\begin{figure}[!t]
\begin{center}
\includegraphics[width=0.9\columnwidth,trim={0 10mm 0 10mm},clip]{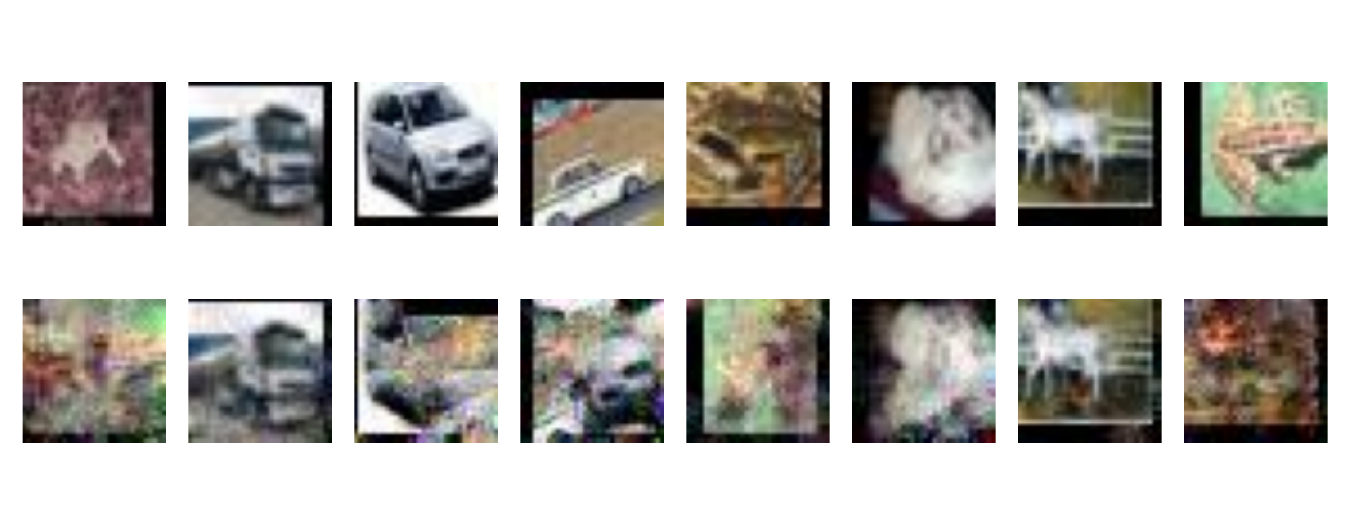}
\end{center}
\vspace*{-5mm}
\caption{\label{fig:ig-bad-example} Reconstructions on CIFAR-10 from Inverting Gradients batch size 8. Duplicate labels exist between images 0, 4, 7 and images 2, 3. Reconstruction 3 contains parts of the content from both images 2 and 3. Reconstructions 0, 4, and 7 appear almost as noise. However, the other images still have reasonable quality.}
\vspace*{-6mm}
\end{figure}

We additionally note a few trends manually observed when going through the GI batch reconstructions of CIFAR-10. 

While some reconstructions have very low quality, only images of the same label will be swapped within the batch during label matching. This is a desirable property for training on leaked data. As long as the set of labels is leaked correctly, the optimization process will match images to a correct label. 
So even though the image quality is poor (and can be entirely unrecognizable), since the label is correctly matched, such an image still does help in the training (this buttresses the message from Tables~\ref{tab:cifar-best-psnr} and~\ref{tab:cifar-worst-psnr}). Naturally, as the PSNR goes down, the image helps less and less. 
% Since some images become very difficult to identify, this will avoid any difficulty with human matching of images since they will be matched with the labels that result in a gradient closest to the ground truth.
% SB (11/17/23): Verify
% JZ (11/17/23): Is correct. The PSNR thing may be specifically because reconstructing images has useful information even if PSNR is low.

% SB (11/17/23): Old text before rewording
\begin{comment}
We also note that when all labels are different within a batch, the reconstruction quality of the entire batch of images is almost always good. An example on CIFAR-10 is shown in Figure~\ref{fig:ig-good-example}. Usually when the reconstructions are poor, it occurs between groups of images in the same class. Poor reconstructions can appear almost as entirely noise, but images with the same label in the batch can occasionally have visible pieces of each image stitched together. These poor reconstructions do not always affect other images within the batch, as those may still be reconstructed well. The example batch in Figure~\ref{fig:ig-bad-example} depicts these properties of poorly reconstructed images. Despite this, we note that the reverse correlation of the same label images and poor reconstructions is not true. Having duplicate labels does not always imply that the reconstructions will be bad. 
\end{comment}
% SB (11/17/23): New text after rewording
We also note that when all labels are different within a batch, the reconstruction quality of the entire batch of images is almost always good. An example on CIFAR-10 is shown in Figure~\ref{fig:ig-good-example}. Usually when the reconstructions are poor, it occurs between groups of images in the same batch that belong to the same class. Poor reconstructions can appear as almost entirely noise, but occasionally images can have visible pieces of each image from the same class in the batch stitched together. The example batch in Figure~\ref{fig:ig-bad-example} depicts these properties of poorly reconstructed images. As a side note, we observe that a batch may have a mix of well reconstructed and poorly reconstructed images. We also note that the reverse mapping  of the same label images to poor reconstructions is not true --- having duplicate labels does not always imply that the reconstructions will be bad.

% \textcolor{red}{Ahaan currently checking for relationship between the cosine similarity of gradients and reconstruction quality. (bad reconstruction - gradients with high cos similarity, but high cos sim doesn't mean bad reconstruction. Correlates with same label observation, as same label has high cos similarity)}

% \subsection{Differential privacy}

% \subsection{Reconstruction quality and cosine similarity}

%% file: sec/5_discussions.tex
\vspace*{-2mm}
\section{Discussion}
\label{sec:discussions}
\vspace*{-2mm}
As a main goal of our work, we have explored the effect of leaked data through today's leading data reconstruction attacks on downstream model training. Our main message is that it is important to also consider the use of data beyond reconstruction quality and image similarity. It is important to consider how far these leaked samples help in a downstream training task. From the experiments, we have seen that current reconstruction attacks are powerful, but still lack in several critical areas. For gradient inversion attacks, the reconstruction quality does pose an issue for training models. Another impediment to practical use of this attack type is the time and computational cost. 
% However, another big challenge lies in computational cost, as the attacks take considerable amounts of time to run even on a powerful GPU like a NVIDIA A100. When compared with linear layer leakage attacks which take a nearly negligible amount of time to reconstruct data, the downside is very apparent.

Linear layer leakage attacks do not suffer from reconstruction quality issues and are much lighter weight computationally, but still suffer from other challenges. While LOKI has increased the efficiency of the FC layer leakage in FedAvg, there is still room for improvement (especially in FedSGD). As shown in~\cite{zhao2023resource}, the model size overhead added by these attacks can be very large, especially if secure aggregation is used. Furthermore, the lack of labels poses a large challenge in training models on the data. With a small dataset like CIFAR-10, hand labeling 40 images is not too challenging. However, larger datasets will require more total labels. For example, SimMatch with $1\%$ of labels on ImageNet can achieve $67.2\%$ top-1 accuracy~\cite{zheng2022simmatch}. However, even $1\%$ of the total images in ImageNet-1k is 12,812 images. Even if the images were leaked, hand labeling would be extremely hard. Furthermore, we see from Figure~\ref{fig:ssl-40} that having less total data and fewer labels results in an even steeper decrease in performance. % Finally, we show that it is even possible to improve federated learning models by centrally training with a small set of leaked data. 
\joshua{Not sure if we will include previous sentence.} \saurabh{Verify} \joshua{Commented out since it was out of place. This part is talking about weaknesses of LLL.}

While not discussed previously, there are many other aspects that affect data reconstruction attacks when considering training models. For example, how does the non-IID aspect of client data affect reconstruction? Other relevant settings of federated learning such as asynchronous federated learning, client selection, and differential privacy can also affect the final leaked dataset. Centralized training also allows for more complex and larger models without the communication or computational restrictions of clients in federated learning. These topics can be explored in future work.

%% file: sec/6_conclusions.tex
\vspace*{-6mm}
\section{Conclusion}
\label{sec:conclusions}
\vspace*{-2mm}
We examine data reconstruction attacks through the lens of training models on the leaked data. While highlighting how the weaknesses of gradient inversion in terms of reconstruction quality affect downstream model training, we also show that even poorly reconstructed images are useful for training. We also discuss how the label matching problem for linear layer leakage can be mitigated through semi-supervised learning. Under ideal conditions, we demonstrate that leaked data from both gradient inversion and linear layer leakage attacks are able to train powerful models comparable to even a centralized baseline. On CIFAR-10, gradient inversion and linear layer leakage attacks achieve $90.34\%$ and $93.16\%$ testing accuracy respectively, $17.58\%$ and $20.40\%$ higher than federated learning and only $4.04\%$ and $1.22\%$ lower compared to centralized training. 

%% file: sec/7_acknowledgements.tex
\vspace*{-1mm}
{\small\noindent \textbf{Acknowledgements.} This material is based in part upon work supported by Adobe Research, the Army Research Office under Contract number W911NF-2020-221, and the National Science Foundation under Grant Number CCF-1919197. Any opinions, findings, and conclusions or recommendations expressed in this material are those of the authors and do not necessarily reflect the views of the sponsors.}

%% file: sec/X_suppl.tex
\clearpage
\setcounter{page}{1}
\maketitlesupplementary
\section{Additional federated learning results}

\begin{table}[]
\scriptsize
\begin{center}
\begin{tabular}{|lccc|}
\hline
\multicolumn{4}{|c|}{CIFAR-10}                                                                                                                                                                                                                                  \\ \hline
\multicolumn{1}{|l|}{}       & \multicolumn{1}{c|}{\begin{tabular}[c]{@{}c@{}}Number \\ of clients\end{tabular}} & \multicolumn{1}{c|}{\begin{tabular}[c]{@{}c@{}}IID (I) or \\ Non-IID (N)\end{tabular}} & \begin{tabular}[c]{@{}c@{}}Test\\ Acc.\end{tabular} \\ \hline
\multicolumn{1}{|l|}{FedAvg} & \multicolumn{1}{c|}{10}                                                           & \multicolumn{1}{c|}{I}                                                                 & \textbf{75.13}                                               \\
\multicolumn{1}{|l|}{}       & \multicolumn{1}{c|}{10}                                                           & \multicolumn{1}{c|}{N}                                                                 & 72.76                                               \\
\multicolumn{1}{|l|}{}       & \multicolumn{1}{c|}{50}                                                           & \multicolumn{1}{c|}{I}                                                                 & 71.45                                               \\
\multicolumn{1}{|l|}{}       & \multicolumn{1}{c|}{50}                                                           & \multicolumn{1}{c|}{N}                                                                 & 68.71                                               \\ \hline
\multicolumn{1}{|l|}{FedSGD} & \multicolumn{1}{c|}{10}                                                           & \multicolumn{1}{c|}{I}                                                                 & 71.24                                               \\
\multicolumn{1}{|l|}{}       & \multicolumn{1}{c|}{10}                                                           & \multicolumn{1}{c|}{N}                                                                 & 68.78                                               \\
\multicolumn{1}{|l|}{}       & \multicolumn{1}{c|}{50}                                                           & \multicolumn{1}{c|}{I}                                                                 & 65.95                                               \\
\multicolumn{1}{|l|}{}       & \multicolumn{1}{c|}{50}                                                           & \multicolumn{1}{c|}{N}                                                                 & 60.88                                               \\ \hline
\multicolumn{4}{|c|}{MNIST}                                                                                                                                                                                                                                     \\ \hline
\multicolumn{1}{|l|}{}       & \multicolumn{1}{c|}{\begin{tabular}[c]{@{}c@{}}Number \\ of clients\end{tabular}} & \multicolumn{1}{c|}{\begin{tabular}[c]{@{}c@{}}IID (I) or \\ Non-IID (N)\end{tabular}} & \begin{tabular}[c]{@{}c@{}}Test\\ Acc.\end{tabular} \\ \hline
\multicolumn{1}{|l|}{FedAvg} & \multicolumn{1}{c|}{10}                                                           & \multicolumn{1}{c|}{I}                                                                 & 96.62                                               \\
\multicolumn{1}{|l|}{}       & \multicolumn{1}{c|}{10}                                                           & \multicolumn{1}{c|}{N}                                                                 & 96.17                                               \\
\multicolumn{1}{|l|}{}       & \multicolumn{1}{c|}{50}                                                           & \multicolumn{1}{c|}{I}                                                                 & 96.68                                               \\
\multicolumn{1}{|l|}{}       & \multicolumn{1}{c|}{50}                                                           & \multicolumn{1}{c|}{N}                                                                 & 96.18                                               \\ \hline
\multicolumn{1}{|l|}{FedSGD} & \multicolumn{1}{c|}{10}                                                           & \multicolumn{1}{c|}{I}                                                                 & 96.68                                               \\
\multicolumn{1}{|l|}{}       & \multicolumn{1}{c|}{10}                                                           & \multicolumn{1}{c|}{N}                                                                 & 96.76                                               \\
\multicolumn{1}{|l|}{}       & \multicolumn{1}{c|}{50}                                                           & \multicolumn{1}{c|}{I}                                                                 & \textbf{96.84}                                               \\
\multicolumn{1}{|l|}{}       & \multicolumn{1}{c|}{50}                                                           & \multicolumn{1}{c|}{N}                                                                 & 96.83                                               \\ \hline
\multicolumn{4}{|c|}{Tiny ImageNet}                                                                                                                                                                                                                             \\ \hline
\multicolumn{1}{|l|}{}       & \multicolumn{1}{c|}{\begin{tabular}[c]{@{}c@{}}Number \\ of clients\end{tabular}} & \multicolumn{1}{c|}{\begin{tabular}[c]{@{}c@{}}IID (I) or \\ Non-IID (N)\end{tabular}} & \begin{tabular}[c]{@{}c@{}}Test\\ Acc.\end{tabular} \\ \hline
\multicolumn{1}{|l|}{FedAvg} & \multicolumn{1}{c|}{10}                                                           & \multicolumn{1}{c|}{I}                                                                 & 37.18                                               \\
\multicolumn{1}{|l|}{}       & \multicolumn{1}{c|}{10}                                                           & \multicolumn{1}{c|}{N}                                                                 & 37.00                                               \\
\multicolumn{1}{|l|}{}       & \multicolumn{1}{c|}{50}                                                           & \multicolumn{1}{c|}{I}                                                                 & \textbf{38.84}                                               \\
\multicolumn{1}{|l|}{}       & \multicolumn{1}{c|}{50}                                                           & \multicolumn{1}{c|}{N}                                                                 & 35.06                                               \\ \hline
\multicolumn{1}{|l|}{FedSGD} & \multicolumn{1}{c|}{10}                                                           & \multicolumn{1}{c|}{I}                                                                 & 35.56                                               \\
\multicolumn{1}{|l|}{}       & \multicolumn{1}{c|}{10}                                                           & \multicolumn{1}{c|}{N}                                                                 & 34.27                                               \\
\multicolumn{1}{|l|}{}       & \multicolumn{1}{c|}{50}                                                           & \multicolumn{1}{c|}{I}                                                                 & 32.77                                               \\
\multicolumn{1}{|l|}{}       & \multicolumn{1}{c|}{50}                                                           & \multicolumn{1}{c|}{N}                                                                 & 26.56                                               \\ \hline
\end{tabular}
\end{center}
\vspace*{-5mm}
\caption{\label{tab:supplement-fl-results} Federated learning test accuracy on CIFAR-10, MNIST, and Tiny ImageNet. A bias of 0.5 is used for the non-IID training. The same settings are used between FedSGD and FedAvg outside of the number of rounds. The number of rounds in FedSGD is $3\times$ the number of rounds in FedAvg (3 local iterations in FedAvg).}
\vspace*{-3mm}
\end{table}

Table~\ref{tab:supplement-fl-results} shows additional test accuracy in federated learning on CIFAR-10, MNIST, and Tiny ImageNet. We include results for both IID and non-IID (with bias$=0.5$). For FedSGD training, we use $3\times$ the number of rounds compared to FedAvg (so the models have seen the same amount data in both cases, as we have 3 local iterations in FedAvg). All other settings are the same. An (expected) observed trend is that IID training outperforms non-IID. Both CIFAR-10 and Tiny ImageNet in FedAvg perform better than FedSGD in all settings. For MNIST, the performance is similar regardless of FedAvg or FedSGD,  IID or non-IID, achieving around $96\%$ accuracy across the board.

\section{Sample reconstructions}
\begin{figure}[]
\vspace*{-1mm}
\begin{center}
\includegraphics[width=0.9\columnwidth]{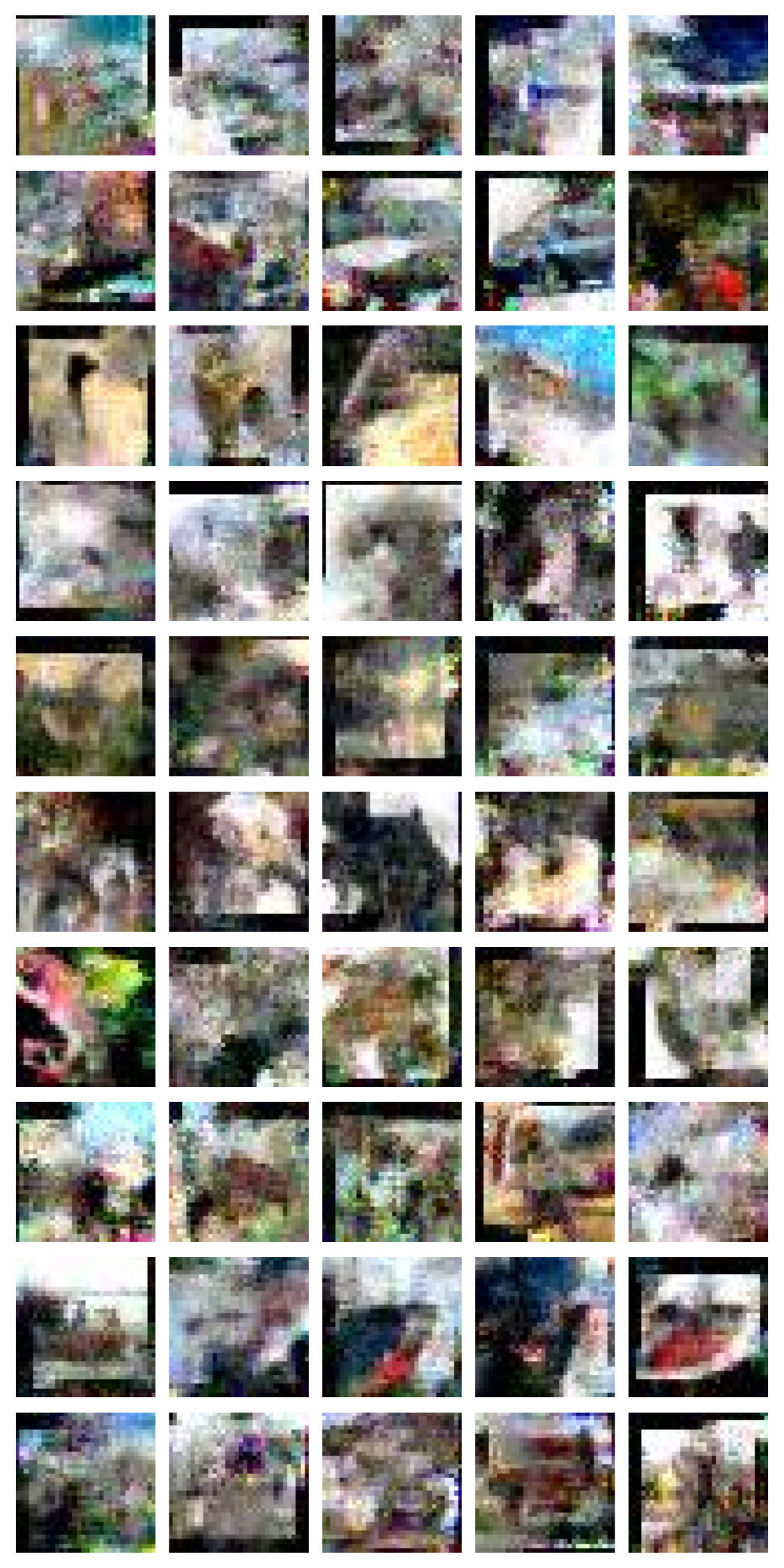}
\end{center}
\vspace*{-5mm}
\caption{\label{fig:supplement-psnr12} CIFAR-10 sample reconstructions from Inverting Gradients batch size 16 with a PSNR $<12$. Each row is a different class. While the images are very noisy, using a set of them for training achieves a model with $45.05\%$ accuracy.}
\vspace*{-5mm}
\end{figure}

\begin{figure}[]
\begin{center}
\includegraphics[width=1.0\columnwidth,trim={20mm 20mm 20mm 20mm},clip]{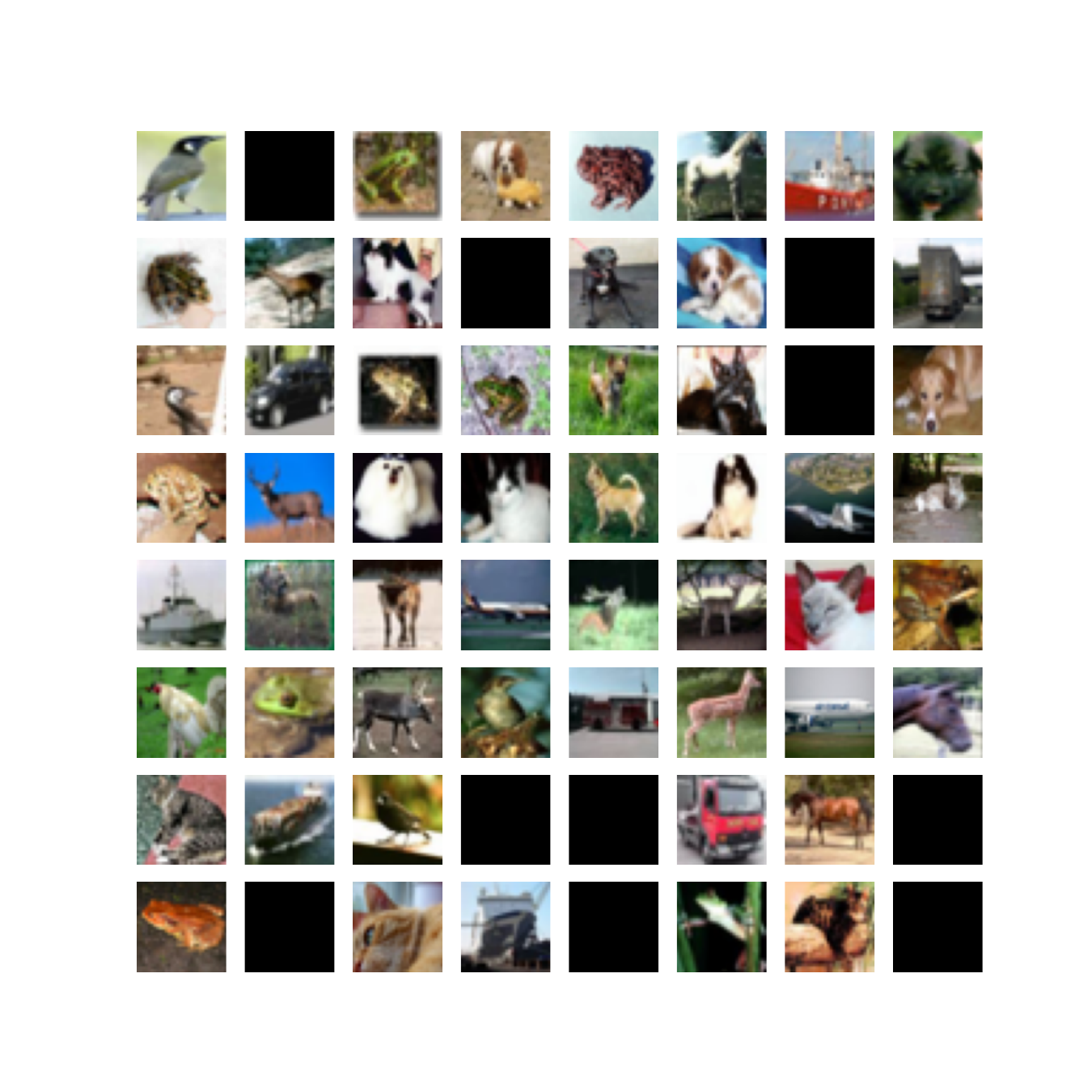}
\end{center}
\vspace*{-5mm}
\caption{\label{fig:supplement-loki-recon} LOKI CIFAR-10 reconstructions using CSF$=500$ in FedAvg. 54 images out of 64 images are leaked.}
\vspace*{-2mm}
\end{figure}

\begin{figure}[]
\begin{center}
\includegraphics[width=1.0\columnwidth,trim={20mm 20mm 20mm 20mm},clip]{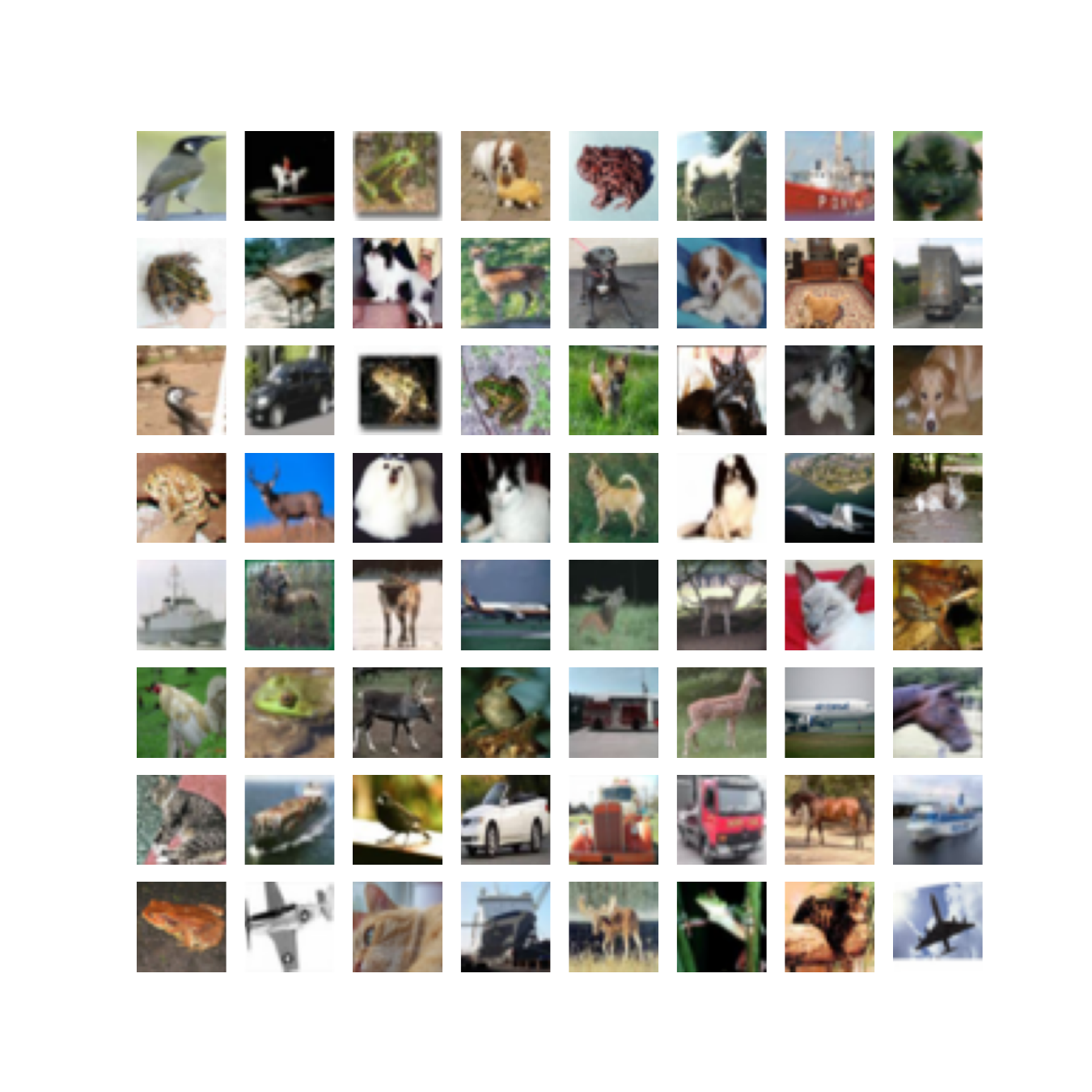}
\end{center}
\vspace*{-5mm}
\caption{\label{fig:supplement-loki-gt} Ground truth CIFAR-10 images.}
\vspace*{-5mm}
\end{figure}

For Inverting Gradients, we use a learning rate of $=0.01$ and total variation of 1e-6 on CIFAR-10. For MNIST, we use a learning rate of $=0.01$ and a total variation of 0. These parameters achieved the best reconstruction quality for us.

Figure~\ref{fig:supplement-psnr12} shows sample reconstructions from Inverting Gradients~\cite{geiping2020inverting} batch size 16 with PSNR $<12$. Each row shows 5 images from each of the classes in CIFAR-10. The rows correspond to airplanes, cars, birds, cats, deer, dogs, frogs, horses, ships, and trucks respectively. While the images are noisy, there is some contextual information that can still be observed in the reconstructions. As discussed in Section~\ref{sec:psnr-quality}, removing these images from the training set results in a small decrease in model performance from $76.83\%$ to $76.16\%$. Training on only the set of images with PSNR $<12$ results in a $45.05\%$ test accuracy.

Figure ~\ref{fig:supplement-loki-recon} shows CIFAR-10 reconstructions from LOKI in FedAvg using CSF$=500$. Figure~\ref{fig:supplement-loki-gt} shows the corresponding ground truth images. For this particular set of images, 54 images out of 64 are leaked ($84.38\%$ leakage rate).

\section{Linear layer leakage method comparison}
\begin{table}[]
\scriptsize
\begin{center}
\begin{tabular}{|l|c|c|c|}
\hline
                                                                           & \begin{tabular}[c]{@{}c@{}}FC size\\ factor\end{tabular} & \begin{tabular}[c]{@{}c@{}}Leakage\\ rate\end{tabular} & \begin{tabular}[c]{@{}c@{}}Test\\ accuracy\end{tabular} \\ \hline
\multirow{3}{*}{LOKI}                                                      & 8                                                        & 87.58                                                  & \textbf{93.16}                                                   \\
                                                                           & 4                                                        & 78.93                                                  & 92.94                                                   \\
                                                                           & 2                                                        & 59.76                                                  & 91.90                                                   \\ \hline
\multirow{3}{*}{\begin{tabular}[c]{@{}l@{}}Robbing\\ the Fed\end{tabular}} & 8                                                        & 87.50                                                  & 93.10                                                   \\
                                                                           & 4                                                        & 78.97                                                  & \textbf{93.02}                                                   \\
                                                                           & 2                                                        & 59.72                                                  & \textbf{92.12}                                                   \\ \hline
\multirow{3}{*}{\begin{tabular}[c]{@{}l@{}}Trap\\ Weights\end{tabular}}    & 8                                                        & 58.11                                                  & 91.84                                                   \\
                                                                           & 4                                                        & 45.92                                                  & 90.09                                                   \\
                                                                           & 2                                                        & 30.46                                                  & 86.38                                                   \\ \hline
\end{tabular}
\end{center}
\vspace*{-5mm}
\caption{\label{tab:supplement-lll-results} Leakage rate and test accuracy on CIFAR-10 for LOKI, Robbing the Fed, and Trap Weights in FedSGD. FC layer size factors of 8, 4, and 2 used with a batch size of 64. Models trained from scratch on leaked data.}
\vspace*{-3mm}
\end{table}

Table~\ref{tab:supplement-lll-results} shows the leakage rate and test accuracy on CIFAR-10 for LOKI~\cite{zhao2023loki}, Robbing the Fed~\cite{fowl2022robbing}, and trap weights~\cite{boenisch2021curious}. Attacks are done in FedSGD with a batch size of 64. LOKI and Robbing the Fed have no additional parameters besides the FC size factor (FC layer size = FC size factor$\times$batch size). For trap weights, in addition to the FC size factor, a scaling factor of 0.96 achieves the highest leakage rate for each FC size factor (checked by 0.1 increments). LOKI and Robbing the Fed achieve very similar leakage rates and model performances. Trap weights has lower leakage rate than both other methods and, as a result, lower model performance for the same FC size factors.

\section{SSIM threshold}
\begin{table}
  \centering
  \begin{subtable}{0.49\linewidth}
    \scriptsize
    \begin{center}
    \begin{tabular}{|l|c|c|}
    \hline
    SSIM             & \begin{tabular}[c]{@{}c@{}}\% imgs \\ kept\end{tabular} & \begin{tabular}[c]{@{}c@{}}Test \\ accuracy\end{tabular} \\ \hline
    $>0.7$ & 17.25                                                     & 72.32                                                    \\
    $>0.6$ & 30.26                                                     & 75.68                                                    \\
    $>0.5$ & 44.76                                                     & 75.94                                                    \\
    $>0.4$ & 61.71                                                     & 76.61                                                    \\
    $>0.3$ & 80.56                                                     &    77.02                                                 \\
    $>0.2$ & 95.55                                                     & 77.31                                                    \\ \hline
    \end{tabular}
    \end{center}
    \vspace*{-1mm}
    \caption{PSNR above threshold}
    \label{tab:supplement-good-ssim}
  \end{subtable}
  \hfill
    \begin{subtable}{0.49\linewidth}
    \scriptsize
    \begin{center}
    \begin{tabular}{|l|c|c|}
    \hline
    SSIM             & \begin{tabular}[c]{@{}c@{}}\% imgs\\ kept\end{tabular} & \begin{tabular}[c]{@{}c@{}}Test \\ accuracy\end{tabular} \\ \hline
    $<0.7$ & 82.75                                                     & 70.94                                                    \\
    $<0.6$ & 69.74                                                     & 61.66                                                    \\
    $<0.5$ & 55.24                                                     & 57.34                                                    \\
    $<0.4$ & 38.29                                                     & 51.56                                                    \\
    $<0.3$ & 19.44                                                     & 47.00                                                    \\
    $<0.2$ & 4.45                                                     & 32.36                                                    \\ \hline
    \end{tabular}
    \end{center}
    \vspace*{-1mm}
    \caption{PSNR below threshold}
    \label{tab:supplement-bad-ssim}
  \end{subtable}
  \vspace*{-3mm}
  \caption{\label{tab:supplement-ssim} Training models using CIFAR-10 leaked data from inverting gradients batch size 16. Only the reconstructions with an SSIM (a) above and (b) below the threshold are used in training.}
  \vspace*{-5mm}
\end{table}

Table~\ref{tab:supplement-ssim} shows the test accuracy of models trained while removing images based on the SSIM. Table~\ref{tab:supplement-good-ssim} shows accuracy when only images \textit{above} an SSIM threshold are used. Table~\ref{tab:supplement-bad-ssim} shows accuracy when images \textit{below} an SSIM threshold are used. For SSIM, removing a set of the worst images with SSIM $<0.2$ or $<0.3$ results in a small model performance increase compared to when all images are included (which achieves $76.83\%$). Similar to PSNR, training on a set of the worst quality reconstructions (SSIM $<0.2$) achieves $32.36\%$ accuracy, a higher accuracy than random guessing, but much lower performance compared to the baseline.

\iffalse
\section{Rationale}
\label{sec:rationale}
% 
Having the supplementary compiled together with the main paper means that:
% 
\begin{itemize}
\item The supplementary can back-reference sections of the main paper, for example, we can refer to \cref{sec:intro};
\item The main paper can forward reference sub-sections within the supplementary explicitly (e.g. referring to a particular experiment); 
\item When submitted to arXiv, the supplementary will already included at the end of the paper.
\end{itemize}
% 
To split the supplementary pages from the main paper, you can use \href{https://support.apple.com/en-ca/guide/preview/prvw11793/mac#:~:text=Delete%20a%20page%20from%20a,or%20choose%20Edit%20%3E%20Delete).}{Preview (on macOS)}, \href{https://www.adobe.com/acrobat/how-to/delete-pages-from-pdf.html#:~:text=Choose%20%E2%80%9CTools%E2%80%9D%20%3E%20%E2%80%9COrganize,or%20pages%20from%20the%20file.}{Adobe Acrobat} (on all OSs), as well as \href{https://superuser.com/questions/517986/is-it-possible-to-delete-some-pages-of-a-pdf-document}{command line tools}.
\fi

%% file: main.bbl
\begin{thebibliography}{32}
\providecommand{\natexlab}[1]{#1}
\providecommand{\url}[1]{\texttt{#1}}
\expandafter\ifx\csname urlstyle\endcsname\relax
  \providecommand{\doi}[1]{doi: #1}\else
  \providecommand{\doi}{doi: \begingroup \urlstyle{rm}\Url}\fi

\bibitem[Boenisch et~al.(2023)Boenisch, Dziedzic, Schuster, Shamsabadi, Shumailov, and Papernot]{boenisch2021curious}
Franziska Boenisch, Adam Dziedzic, Roei Schuster, Ali~Shahin Shamsabadi, Ilia Shumailov, and Nicolas Papernot.
\newblock When the curious abandon honesty: Federated learning is not private.
\newblock \emph{8th IEEE European Symposium on Security and Privacy (IEEE Euro S\&P)}, 2023.

\bibitem[Bonawitz et~al.(2017)Bonawitz, Ivanov, Kreuter, Marcedone, McMahan, Patel, Ramage, Segal, and Seth]{bonawitz2017practical}
Keith Bonawitz, Vladimir Ivanov, Ben Kreuter, Antonio Marcedone, H~Brendan McMahan, Sarvar Patel, Daniel Ramage, Aaron Segal, and Karn Seth.
\newblock Practical secure aggregation for privacy-preserving machine learning.
\newblock In \emph{proceedings of the 2017 ACM SIGSAC Conference on Computer and Communications Security}, pages 1175--1191, 2017.

\bibitem[Choquette-Choo et~al.(2021)Choquette-Choo, Tramer, Carlini, and Papernot]{choquette2021label}
Christopher~A Choquette-Choo, Florian Tramer, Nicholas Carlini, and Nicolas Papernot.
\newblock Label-only membership inference attacks.
\newblock In \emph{International conference on machine learning}, pages 1964--1974. PMLR, 2021.

\bibitem[Fan et~al.(2020)Fan, Ng, Ju, Zhang, Liu, Chan, and Yang]{fan2020rethinking}
Lixin Fan, Kam~Woh Ng, Ce Ju, Tianyu Zhang, Chang Liu, Chee~Seng Chan, and Qiang Yang.
\newblock Rethinking privacy preserving deep learning: How to evaluate and thwart privacy attacks.
\newblock In \emph{Federated Learning}, pages 32--50. Springer, 2020.

\bibitem[Fowl et~al.(2022)Fowl, Geiping, Czaja, Goldblum, and Goldstein]{fowl2022robbing}
Liam~H Fowl, Jonas Geiping, Wojciech Czaja, Micah Goldblum, and Tom Goldstein.
\newblock Robbing the fed: Directly obtaining private data in federated learning with modified models.
\newblock In \emph{International Conference on Learning Representations}, 2022.

\bibitem[Geiping et~al.(2020)Geiping, Bauermeister, Dr{\"o}ge, and Moeller]{geiping2020inverting}
Jonas Geiping, Hartmut Bauermeister, Hannah Dr{\"o}ge, and Michael Moeller.
\newblock Inverting gradients-how easy is it to break privacy in federated learning?
\newblock \emph{Advances in Neural Information Processing Systems}, 33:\penalty0 16937--16947, 2020.

\bibitem[Geng et~al.(2021)Geng, Mou, Li, Li, Beyan, Decker, and Rong]{geng2021towards}
Jiahui Geng, Yongli Mou, Feifei Li, Qing Li, Oya Beyan, Stefan Decker, and Chunming Rong.
\newblock Towards general deep leakage in federated learning.
\newblock \emph{arXiv preprint arXiv:2110.09074}, 2021.

\bibitem[He et~al.(2016)He, Zhang, Ren, and Sun]{he2016deep}
Kaiming He, Xiangyu Zhang, Shaoqing Ren, and Jian Sun.
\newblock Deep residual learning for image recognition.
\newblock In \emph{Proceedings of the IEEE conference on computer vision and pattern recognition}, pages 770--778, 2016.

\bibitem[Hitaj et~al.(2017)Hitaj, Ateniese, and Perez-Cruz]{hitaj2017deep}
Briland Hitaj, Giuseppe Ateniese, and Fernando Perez-Cruz.
\newblock Deep models under the gan: information leakage from collaborative deep learning.
\newblock In \emph{Proceedings of the 2017 ACM SIGSAC conference on computer and communications security}, pages 603--618, 2017.

\bibitem[Huang et~al.(2021)Huang, Gupta, Song, Li, and Arora]{huang2021evaluating}
Yangsibo Huang, Samyak Gupta, Zhao Song, Kai Li, and Sanjeev Arora.
\newblock Evaluating gradient inversion attacks and defenses in federated learning.
\newblock \emph{Advances in Neural Information Processing Systems}, 34:\penalty0 7232--7241, 2021.

\bibitem[Krizhevsky et~al.(2009)Krizhevsky, Hinton, et~al.]{krizhevsky2009learning}
Alex Krizhevsky, Geoffrey Hinton, et~al.
\newblock Learning multiple layers of features from tiny images.
\newblock Master's thesis, University of Toronto, 2009.

\bibitem[Le and Yang(2015)]{le2015tiny}
Ya Le and Xuan Yang.
\newblock Tiny imagenet visual recognition challenge.
\newblock \emph{CS 231N}, 7\penalty0 (7):\penalty0 3, 2015.

\bibitem[LeCun(1998)]{lecun1998mnist}
Yann LeCun.
\newblock The mnist database of handwritten digits.
\newblock \emph{http://yann. lecun. com/exdb/mnist/}, 1998.

\bibitem[Li et~al.(2021)Li, Xiong, and Hoi]{li2021comatch}
Junnan Li, Caiming Xiong, and Steven~CH Hoi.
\newblock Comatch: Semi-supervised learning with contrastive graph regularization.
\newblock In \emph{Proceedings of the IEEE/CVF International Conference on Computer Vision}, pages 9475--9484, 2021.

\bibitem[Luo et~al.(2021)Luo, Wu, Xiao, and Ooi]{luo2021feature}
Xinjian Luo, Yuncheng Wu, Xiaokui Xiao, and Beng~Chin Ooi.
\newblock Feature inference attack on model predictions in vertical federated learning.
\newblock In \emph{2021 IEEE 37th International Conference on Data Engineering (ICDE)}, pages 181--192. IEEE, 2021.

\bibitem[Ma et~al.(2022)Ma, Sun, Cui, Li, Guan, and Liu]{ma2022instance}
Kailang Ma, Yu Sun, Jian Cui, Dawei Li, Zhenyu Guan, and Jianwei Liu.
\newblock Instance-wise batch label restoration via gradients in federated learning.
\newblock In \emph{The Eleventh International Conference on Learning Representations}, 2022.

\bibitem[McMahan et~al.(2017)McMahan, Moore, Ramage, Hampson, and y~Arcas]{mcmahan2017communication}
Brendan McMahan, Eider Moore, Daniel Ramage, Seth Hampson, and Blaise~Aguera y Arcas.
\newblock Communication-efficient learning of deep networks from decentralized data.
\newblock In \emph{Artificial intelligence and statistics}, pages 1273--1282. PMLR, 2017.

\bibitem[Melis et~al.(2019)Melis, Song, De~Cristofaro, and Shmatikov]{melis2019exploiting}
Luca Melis, Congzheng Song, Emiliano De~Cristofaro, and Vitaly Shmatikov.
\newblock Exploiting unintended feature leakage in collaborative learning.
\newblock In \emph{2019 IEEE symposium on security and privacy (SP)}, pages 691--706. IEEE, 2019.

\bibitem[Nasr et~al.(2019)Nasr, Shokri, and Houmansadr]{nasr2019comprehensive}
Milad Nasr, Reza Shokri, and Amir Houmansadr.
\newblock Comprehensive privacy analysis of deep learning: Passive and active white-box inference attacks against centralized and federated learning.
\newblock In \emph{2019 IEEE symposium on security and privacy (SP)}, pages 739--753. IEEE, 2019.

\bibitem[Pasquini et~al.(2022)Pasquini, Francati, and Ateniese]{pasquini2021eluding}
Dario Pasquini, Danilo Francati, and Giuseppe Ateniese.
\newblock Eluding secure aggregation in federated learning via model inconsistency.
\newblock In \emph{Proceedings of the 2022 ACM SIGSAC Conference on Computer and Communications Security}, pages 2429--2443, 2022.

\bibitem[Phong et~al.(2017)Phong, Aono, Hayashi, Wang, and Moriai]{phong2017privacy}
Le~Trieu Phong, Yoshinori Aono, Takuya Hayashi, Lihua Wang, and Shiho Moriai.
\newblock Privacy-preserving deep learning: Revisited and enhanced.
\newblock In \emph{International Conference on Applications and Techniques in Information Security}, pages 100--110. Springer, 2017.

\bibitem[Shokri et~al.(2017)Shokri, Stronati, Song, and Shmatikov]{shokri2017membership}
Reza Shokri, Marco Stronati, Congzheng Song, and Vitaly Shmatikov.
\newblock Membership inference attacks against machine learning models.
\newblock In \emph{2017 IEEE symposium on security and privacy (SP)}, pages 3--18. IEEE, 2017.

\bibitem[Wang et~al.(2019)Wang, Song, Zhang, Song, Wang, and Qi]{wang2019beyond}
Zhibo Wang, Mengkai Song, Zhifei Zhang, Yang Song, Qian Wang, and Hairong Qi.
\newblock Beyond inferring class representatives: User-level privacy leakage from federated learning.
\newblock In \emph{IEEE INFOCOM 2019-IEEE Conference on Computer Communications}, pages 2512--2520. IEEE, 2019.

\bibitem[Wen et~al.(2022)Wen, Geiping, Fowl, Goldblum, and Goldstein]{wen2022fishing}
Yuxin Wen, Jonas Geiping, Liam Fowl, Micah Goldblum, and Tom Goldstein.
\newblock Fishing for user data in large-batch federated learning via gradient magnification.
\newblock \emph{International Conference on Machine Learning}, 2022.

\bibitem[Yang et~al.(2022)Yang, Wu, Zhang, Jiang, Liu, Zheng, Zhang, Wang, and Zeng]{yang2022class}
Fan Yang, Kai Wu, Shuyi Zhang, Guannan Jiang, Yong Liu, Feng Zheng, Wei Zhang, Chengjie Wang, and Long Zeng.
\newblock Class-aware contrastive semi-supervised learning.
\newblock In \emph{Proceedings of the IEEE/CVF Conference on Computer Vision and Pattern Recognition}, pages 14421--14430, 2022.

\bibitem[Yin et~al.(2021)Yin, Mallya, Vahdat, Alvarez, Kautz, and Molchanov]{yin2021see}
Hongxu Yin, Arun Mallya, Arash Vahdat, Jose~M Alvarez, Jan Kautz, and Pavlo Molchanov.
\newblock See through gradients: Image batch recovery via gradinversion.
\newblock In \emph{Proceedings of the IEEE/CVF Conference on Computer Vision and Pattern Recognition}, pages 16337--16346, 2021.

\bibitem[Zagoruyko and Komodakis(2016)]{zagoruyko2016wide}
Sergey Zagoruyko and Nikos Komodakis.
\newblock Wide residual networks.
\newblock \emph{arXiv preprint arXiv:1605.07146}, 2016.

\bibitem[Zhao et~al.(2020)Zhao, Mopuri, and Bilen]{zhao2020idlg}
Bo Zhao, Konda~Reddy Mopuri, and Hakan Bilen.
\newblock idlg: Improved deep leakage from gradients.
\newblock \emph{arXiv preprint arXiv:2001.02610}, 2020.

\bibitem[Zhao et~al.(2023{\natexlab{a}})Zhao, Elkordy, Sharma, Ezzeldin, Avestimehr, and Bagchi]{zhao2023resource}
Joshua~C Zhao, Ahmed~Roushdy Elkordy, Atul Sharma, Yahya~H Ezzeldin, Salman Avestimehr, and Saurabh Bagchi.
\newblock The resource problem of using linear layer leakage attack in federated learning.
\newblock In \emph{Proceedings of the IEEE/CVF Conference on Computer Vision and Pattern Recognition}, pages 3974--3983, 2023{\natexlab{a}}.

\bibitem[Zhao et~al.(2023{\natexlab{b}})Zhao, Sharma, Elkordy, Ezzeldin, Avestimehr, and Bagchi]{zhao2023loki}
Joshua~Christian Zhao, Atul Sharma, Ahmed~Roushdy Elkordy, Yahya~H Ezzeldin, Salman Avestimehr, and Saurabh Bagchi.
\newblock Loki: Large-scale data reconstruction attack against federated learning through model manipulation.
\newblock In \emph{2024 IEEE Symposium on Security and Privacy (SP)}, pages 30--30. IEEE Computer Society, 2023{\natexlab{b}}.

\bibitem[Zheng et~al.(2022)Zheng, You, Huang, Wang, Qian, and Xu]{zheng2022simmatch}
Mingkai Zheng, Shan You, Lang Huang, Fei Wang, Chen Qian, and Chang Xu.
\newblock Simmatch: Semi-supervised learning with similarity matching.
\newblock In \emph{Proceedings of the IEEE/CVF Conference on Computer Vision and Pattern Recognition}, pages 14471--14481, 2022.

\bibitem[Zhu et~al.(2019)Zhu, Liu, and Han]{zhu2019deep}
Ligeng Zhu, Zhijian Liu, and Song Han.
\newblock Deep leakage from gradients.
\newblock \emph{Advances in neural information processing systems}, 32, 2019.

\end{thebibliography}
